# MACHINE LEARNING-DRIVEN ENZYME MINING: OPPORTUNITIES, CHALLENGES, AND FUTURE PERSPECTIVES

**Yanzi Zhang[1†], Felix Moorhoff[2†], Sizhe Qiu[3], Wenjuan Dong[1], David Medina-Ortiz[2,4], Jing Zhao[1*], and Mehdi D. Davari[2*]**

[1]State Key Laboratory of Biocatalysis and Enzyme Engineering, Hubei Key Laboratory of Industrial Biotechnology, School of Life Sciences, Hubei University, Wuhan, 430062, China
[2]Leibniz-Institute of Plant Biochemistry Department of Bioorganic Chemistry, Weinberg 3, D-06120 Halle, Germany
[3]Department of Engineering Science, University of Oxford, OX1 3PJ, United Kingdom
† These authors contributed equally
* Corresponding Author: Jing Zhao (zhaojing@hubu.edu.cn; zhaojing802@hotmail.com)
* Corresponding Author: Mehdi D. Davari (Mehdi.Davari@ipb-halle.de)

## Abstract

Enzyme mining is rapidly evolving as a data-driven strategy to identify biocatalysts with tailored functions from the vast landscape of uncharacterized proteins. The integration of machine learning (ML) into these workflows enables high-throughput prediction of enzyme functions—including Enzyme Commission numbers, Gene Ontology terms, and substrate specificity—as well as key catalytic properties such as kinetic parameters, optimal temperature, pH, solubility, and thermophilicity. This review provides a systematic overview of state-of-the-art ML models and highlights representative case studies that demonstrate their effectiveness in accelerating enzyme discovery. Despite notable progress, current approaches remain limited by data scarcity, model generalizability, and interpretability. We discuss emerging strategies to overcome these challenges, including multi-task learning, integration of multi-modal data, and explainable AI. Together, these developments establish ML-guided enzyme mining as a scalable and predictive framework for uncovering novel biocatalysts, with broad applications in biocatalysis, biotechnology, and synthetic biology.

## 1 Introduction

Enzymes are essential biocatalysts that mediate the intricate chemical transformations that sustain all forms of life (Robinson, 2015). Their remarkable catalytic versatility has made them indispensable in biotechnology, synthetic biology, and industrial biocatalysis (Katz et al., 2018). Despite the relevance of enzymatic sequences catalogued in public databases, a significant percentage of enzymes remain uncharacterized, represented only by primary sequence data without experimentally validated functions (Shinde et al., 2024). Unlocking their biotechnological potential requires robust functional annotation frameworks and predictive models capable of inferring key biochemical properties directly from sequence information (Zhou et al., 2024).

Historically, enzyme discovery has relied on cultivation-based methods, involving the isolation, cultivation, and biochemical screening (Ahmed et al., 2024). While effective within a limited subset of culturable taxa, these methods are constrained by the fact that most microbial species cannot be readily cultivated under laboratory conditions (Kapinusova et al., 2023). The emergence of metagenomics has significantly expanded access to the enzymatic sequence space by enabling the direct sequencing of environmental DNA (Ariaeenejad et al., 2024). Platforms such as MGnify now host millions of protein-coding sequences. However, functional annotation remains sparse, limiting both biological interpretation and translational application (Richardson et al., 2023).

To overcome this bottleneck, enzyme mining has emerged as a computational strategy to bridge sequence data with putative enzymatic function (Scherlach and Hertweck, 2021). Tools such as sequence similarity networks (SSNs) (Oberg et al., 2023), EnzymeMiner (Hon et al., 2020), and comparative genomics enable the prioritization of enzyme candidates based on homology, conserved motifs, and genomic context. These approaches enhance the efficiency of discovery pipelines by facilitating the identification of high-confidence targets for experimental validation.

Beyond the selection of candidate sequences, computational enzyme mining strategies increasingly guide experimental design by generating hypothesis-driven priorities. These prioritized targets enhance the efficiency of the design–build–test–learn cycle, contribute to the refinement of training datasets, and extend the applicability of machine learning (ML) models across diverse enzymatic contexts (Feehan et al., 2021). Collectively, these contributions enhance the integration of *in silico* predictions with empirical validation, thereby accelerating the overall discovery process (Funk et al., 2024).

Despite these advances, significant challenges remain. Many candidate sequences lack standardized functional annotations, such as Enzyme Commission (EC) numbers or Gene Ontology (GO) terms, complicating benchmarking and downstream integration (Sapoval et al., 2022). Concurrently, innovations in next-generation sequencing, microfluidics, and high-throughput screening have increased the efficiency and reduced the cost of experimental assays (Vasina et al., 2020). Yet, validation efforts remain time- and resource-intensive, necessitating scalable computational solutions to keep pace with the deluge of sequence data (Ao et al., 2024).

Recent advances in high-throughput functional screening have yielded expansive datasets that now support a new wave of ML applications in enzyme discovery (Markel et al., 2020). Models trained on these datasets can predict properties such as catalytic activity, substrate specificity, and physicochemical attributes, thus guiding more targeted experimental workflows (Zhou and Huang, 2024). These approaches not only alleviate experimental burden but also deepen mechanistic understanding of sequence–function relationships (Markus et al., 2023).

In this work, we provide a comprehensive examination of ML strategies for enzyme mining, emphasizing their role in accelerating the identification, annotation, and characterization of biocatalysts across diverse functional and physicochemical dimensions. We explore recent advancements in functional classification—including the prediction of EC numbers, GO terms, and substrate specificity—as well as the estimation of key enzymatic properties such as kinetic parameters, thermostability, pH optima, and solubility. Particular attention is given to the architectural diversity of ML models, including supervised, unsupervised, and multimodal frameworks, and their integration into predictive and scalable discovery pipelines. We further highlight representative case studies demonstrating the practical utility of these tools and critically discuss persistent challenges, including data scarcity, limited interpretability, and model generalizability. Finally, we propose a modular, ML-guided enzyme mining strategy and outline future opportunities toward autonomous, closed-loop discovery platforms that couple computational prediction with experimental validation—laying the foundation for next-generation enzyme biotechnology.

## 2 Enzyme mining as a central framework for functional biocatalyst discovery

Enzyme mining has emerged as a pivotal strategy in modern biotechnology, enabling the systematic identification of functional biocatalysts from the vast, largely uncharacterized sequence space encoded in genomic and metagenomic datasets (Ariaeenejad et al., 2024).

In contrast to traditional enzyme discovery approaches—which rely on the biochemical screening of cultured microorganisms and are limited by cultivation biases and phylogenetic redundancy (Robinson et al., 2021; Kapinusova et al., 2023)—enzyme mining bypasses the need for laboratory cultivation, allowing direct access to enzymes from extremophilic or otherwise inaccessible microbial sources (Bauman et al., 2021). This computational paradigm expands the reach of biocatalyst discovery beyond conventional constraints, facilitating large-scale interrogation of microbial diversity to uncover enzymes with novel or atypical catalytic functions. However, its effectiveness depends critically on accurate functional annotation in the absence of experimental validation, and carries the risk of propagating erroneous predictions through homology-based inference (Yang et al., 2024a).

Conceptually, enzyme mining sits at the intersection of enzyme discovery and enzyme engineering. While the former aims to uncover new enzymes through experimental or metagenomic exploration, and the latter focuses on tailoring known enzymes for specific applications via rational design or directed evolution (Medina-Ortiz et al., 2024c; Daud et al., 2025), enzyme mining bridges these domains by computationally prioritizing candidate sequences with desirable attributes for downstream optimization. As such, it plays an increasingly central role in biotechnological workflows, informing both the selection of novel enzymes and their integration into engineering pipelines (Scherlach and Hertweck, 2021).

The conventional enzyme mining process is typically structured into a series of interdependent stages, including: i) the creation of a tailored enzyme pool, ii) sequence-level characterization and diversity analysis, iii) functional annotation and candidate prioritization, and iv) experimental validation of selected targets (Figure **1**) (Scherer et al., 2021). These steps, while effective, face growing limitations in scalability and resolution—particularly as metagenomic repositories continue to expand at an exponential pace. In the following sections, we examine how machine learning is being incorporated into each of these stages, offering predictive, scalable alternatives that overcome key bottlenecks and extend the capabilities of enzyme mining into new functional and evolutionary landscapes.

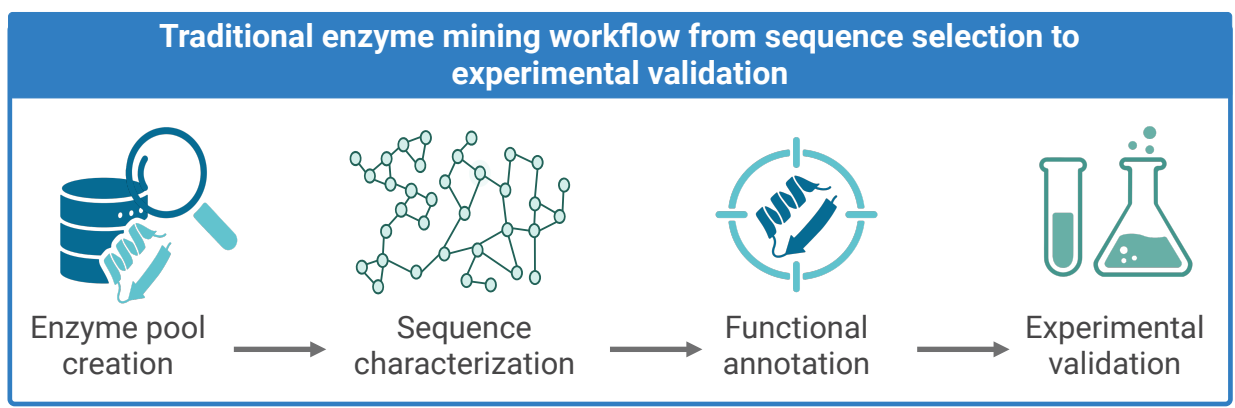

**Figure 1**: **Traditional enzyme mining workflow from sequence selection to experimental validation**. The workflow begins with enzyme pool creation, in which candidate sequences are selected from protein databases based on functional or structural criteria. This is followed by sequence characterization, where sequence diversity and relationships are analyzed using clustering algorithms, sequence similarity networks, or phylogenetic methods. In the third stage, functional annotation involves computational prediction of enzymatic functions and the prioritization of candidates for downstream validation. The final step, experimental validation, confirms predicted functions through biochemical assays and high-throughput screening. Collectively, these steps constitute a cyclical and iterative pipeline for the discovery and characterization of novel enzymes from genomic and metagenomic data.

The process begins with the construction of a tailored enzyme pool, in which protein sequences are selected according to specific research goals—ranging from targeted biocatalyst identification to broader explorations of enzyme superfamilies (Minton, 2001). Sequence retrieval often involves homology searches using BLAST or PSI-BLAST (Altschul et al., 1997), structural comparisons via tools like Foldseek (Van Kempen et al., 2024) or the ESM Metagenomic Atlas (Yeo et al., 2025), and domain-based searches using integrated platforms such as InterPro (Paysan-Lafosse et al., 2023). Automated tools like EnzymeMiner optimize this stage by filtering sequences based on EC numbers and clustering them by functional features (Hon et al., 2020), thereby enhancing the specificity and scalability of pool generation.

Once assembled, the enzyme pool is analyzed through to sequence-level characterization to assess diversity, redundancy, and evolutionary relationships. Sequence similarity networks, phylogenetic trees, and clustering algorithms (e.g., CD-Hit (Fu et al., 2012), DBSCAN (Fuchs and Höpken, 2022), MMSeqs2 (Steinegger and Söding, 2017)) are commonly employed to reduce redundancy and group sequences into functionally coherent clusters. Visualization tools like EFI-EST provide dynamic thresholding for SSN construction (Oberg et al., 2023), enabling the detection of both closely related homologs and distantly related enzymes with potential functional novelty.

The next phase, functional annotation, is the most critical and challenging component of the pipeline. In this step, bioinformatic predictions are used to infer catalytic function, substrate specificity, or environmental tolerance based on sequence motifs, conserved domains, or inferred structural features (Kim et al., 2023). While homology-based approaches and curated databases remain foundational, they often fall short when dealing with remote homologs, promiscuous enzymes, or sequences from poorly characterized taxa (Yang et al., 2024a). These limitations have provoked the growing interest in more sophisticated annotation techniques, including those based on machine learning (Zhou and Huang, 2024), which will be addressed in the next section.

The final stage of the pipeline involves experimental validation, where top-ranking candidates are expressed, purified, and subjected to biochemical assays to confirm predicted functions. Parameters such as catalytic activity, substrate range, thermal stability, and pH tolerance are typically assessed (Gantz et al., 2023). In recent years, high-throughput screening platforms have increased the speed and scale of validation, supporting iterative cycles of discovery and optimization (Longwell et al., 2017). Feedback from experimental results is can be used to refine selection criteria or retrain predictive models, reinforcing the pipeline's adaptive potential.

Despite its utility, the conventional enzyme mining framework faces several persistent challenges. Functional annotation remains heavily dependent on homology inference, which limits discovery in sequence-divergent regions (Boger et al., 2025). Redundancy and data imbalance in public repositories introduce biases that obscure rare but biotechnologically valuable enzymes (Albright and Louca, 2023). Moreover, the increasing volume of metagenomic data surpasses the analytical capacity of many traditional tools, making prioritization and interpretation difficult at scale (Feehan et al., 2021). These limitations underscore the need for scalable, accurate, and generalizable computational frameworks—motivating the integration of machine learning into enzyme mining pipelines.

As will be discussed in the next sections, ML-based approaches offer significant advantages in functional prediction, candidate prioritization, and sequence-to-function generalization. Their integration into enzyme mining is redefining the landscape of biocatalyst discovery, enabling a shift from rule-based screening to data-driven inference at scale.

## 3  Machine learning strategies for enzyme discovery and mining

The advances of high-throughput sequencing technologies and the exponential growth of genomic and metagenomic repositories have significantly expanded the sequence space available for enzyme discovery (Silva et al., 2023). However, extracting functional insight from this vast and heterogeneous landscape demands computational frameworks capable of generalizing beyond homology and conventional rule-based heuristics (Yang et al., 2024a). Machine learning has emerged as a transformative approach in this context, offering data-driven strategies for modeling complex sequence–function relationships, predicting enzymatic properties, and enabling the scalable prioritization of biocatalyst candidates across diverse taxonomic and ecological contexts (Markus et al., 2023).

In this section, we explore the conceptual foundations and methodological advances of ML-based enzyme mining. We begin by reviewing core learning paradigms—including supervised, unsupervised, and generative modeling—as they pertain to enzyme biotechnology. We then describe early data-driven strategies that enabled the way for current ML architectures, highlighting their evolution toward deep learning and protein language modeling. Particular attention is given to predictive frameworks for functional annotation (e.g., EC numbers, GO terms, substrate specificity) and enzymatic property estimation (e.g., kinetics, thermostability, solubility), along with the major challenges that persist in model generalizability, interpretability, and data availability. Lastly, we present representative case studies to illustrate how these approaches are being applied in real-world enzyme mining scenarios, setting the stage for the development of fully autonomous ML-guided discovery pipelines.

### 3.1  Foundations of data-driven modeling and machine learning in enzyme biotechnology

Data-driven strategies have become central to modern computational biology, enabling predictive and exploratory analyses that complement traditional rule-based approaches (Libbrecht and Noble, 2015). In the context of enzyme biotechnology, ML serves as a powerful paradigm for modeling complex sequence–function relationships, navigating high-dimensional data spaces, and generating actionable hypotheses from large-scale biological datasets (Landwehr et al., 2025).

ML comprises several categories of learning paradigms. Supervised learning is widely used to predict functional annotations or biochemical properties based on labeled datasets (Landeta et al., 2024; Medina-Ortiz et al., 2024a). Unsupervised learning supports exploratory tasks such as clustering enzyme families or discovering latent sequence embeddings (David and Halanych, 2023). Reinforcement learning—though less common—has been applied in tasks such as *de novo* enzyme design and optimization (Wang et al., 2025c). Recently, generative learning has gained prominence, enabling the generation of novel enzyme sequences or the completion of partially characterized ones using deep generative models (Goles et al., 2024; Medina-Ortiz et al., 2024b).

The traditional data-driven workflow for ML model development follows a structured sequence of stages. It begins with data collection, which may involve extracting sequences and annotations from public repositories or experimental datasets (Andersen and Reading, 2024). The second step, data representation, is critical and may rely on handcrafted features (feature engineering), classical encoding schemes (e.g., one-hot, k-mer frequency), or more recent embedding techniques derived from pre-trained protein language models (Harding-Larsen et al., 2024). Structural encoders and geometric deep learning (GDL) approaches have also emerged as promising alternatives for capturing spatial and topological information from enzyme structures or graphs (García-Vinuesa et al., 2025).

Following data preparation, model training is performed using either classical algorithms (e.g., support vector machines, random forests) or more complex architectures such as deep neural networks and transformers (Kumar et al., 2023). In many cases, fine-tuning of pre-trained models or integrating GDL-based architectures provides an effective strategy for improving generalization in biological tasks (Medina-Ortiz et al., 2024b; García-Vinuesa et al., 2025). Model performance is then assessed using metrics appropriate to the problem type. For classification tasks commonly rely on accuracy, precision, recall, F1-score, or area under the ROC curve. In contrast, regression tasks use metrics such as mean squared error (MSE), root mean squared error (RMSE), or Pearson coefficient (Medina-Ortiz et al., 2022).

To enhance model performance, hyperparameter optimization strategies such as grid search, random search, genetic algorithms, or Bayesian optimization are often employed (Erden et al., 2023). Finally, trained models may be deployed for inference on new data or integrated into broader pipelines for enzyme discovery, annotation, or design, often using modular or web-based systems for scalability and reproducibility (Alla and Adari, 2020).

Despite these methodological advances, the integration of ML into enzyme biotechnology faces several persistent challenges. Data scarcity remains a fundamental limitation, particularly in so-called low-N regimes where labeled data are limited or imbalanced (Biswas et al., 2021). In many cases, public datasets are biased toward overrepresented enzyme families or well-studied model organisms, leading to annotation bias and reduced generalizability (Radivojac et al., 2013). Furthermore, the interpretability of ML models—especially deep learning architectures—poses a barrier to adoption in biological settings where mechanistic understanding is essential (Medina-Ortiz et al., 2024c). Efforts to incorporate explainable AI (XAI), integrate uncertainty estimation, and prioritize transparency are increasingly necessary to bridge the gap between computational prediction and experimental validation (Teufel et al., 2025).

These foundations define the conceptual and methodological landscape for applying machine learning to enzyme mining and discovery. The following sections give details about specific applications, including functional annotation and the prediction of enzymatic properties.

## 3.2 Early data-driven strategies for enzyme mining

In this section, we describe the early data-driven approaches that laid the foundation for modern machine learning in enzyme mining. These methods enabled the functional classification and property prediction of enzymes using structured datasets and statistical learning, marking a departure from traditional screening workflows that relied on cultivation and experimental throughput constraints (Zhou and Huang, 2024).

We first review computational strategies developed for functional annotation—including EC number assignment, GO term prediction, and substrate specificity classification—followed by models focused on estimating enzymatic properties such as catalytic efficiency, thermostability, solubility, and optimal reaction conditions. Although early models were limited in their capacity to generalize beyond well-characterized enzyme families, they introduced scalable, quantitative frameworks that now underpin modern predictive pipelines.

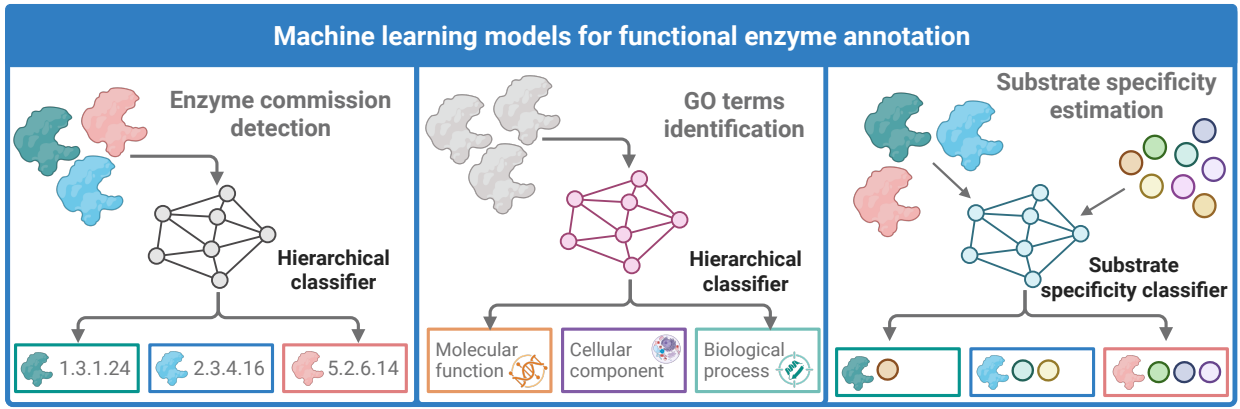

**Figure 2: Machine learning frameworks for functional enzyme annotation.** The figure presents a conceptual overview of machine learning strategies applied to the functional characterization of enzymes from protein sequence data or protein structure. (*Left*) EC number classification models predict enzymatic function across the four-level EC hierarchy, capturing enzyme-specific biochemical activities and substrate-product relationships. (*Center*) Gene Ontology (GO) term prediction integrates molecular function, biological process, and cellular component annotations using deep neural architectures that incorporate structural, evolutionary, and ontological information. (*Right*) Substrate specificity models estimate enzyme–substrate compatibility through binary or multi-class prediction schemes, enabling rational prioritization of candidate enzymes for experimental validation. Collectively, these models constitute the core of modern data-driven enzyme mining pipelines, facilitating scalable, high-throughput, and biologically meaningful functional annotation.

### 3.2.1 Machine learning models for functional enzyme annotation

The functional annotation of enzymes plays a central role in enzyme biotechnology, enabling the identification, classification, and repurposing of biocatalysts for industrial, pharmaceutical, and environmental applications(Ariaeenejad et al., 2024). While traditional approaches based on activity assays, structural analysis, and comparative genomics remain reliable, they are resource-intensive and increasingly unable to keep pace with the rapid expansion of genomic and metagenomic datasets. ML has emerged as a powerful alternative, providing predictive frameworks that infer enzymatic functions directly from sequence and structural features (Han et al., 2024).

ML-based classification models are now integral to enzyme mining workflows. These models can predict enzyme function at multiple levels, including reaction type, catalytic mechanism, and substrate specificity (Wang et al., 2025d). By leveraging diverse data modalities—such as amino acid sequences, structural motifs, and molecular representations—ML approaches have advanced the prediction of EC numbers, GO terms, and enzyme-substrate interactions (Kulmanov et al., 2024).

Advanced ML architectures, including convolutional neural networks (CNNs), graph convolutional networks (GCNs), and transformers, have significantly improved predictive accuracy across diverse enzyme families (Mao et al., 2025). CNNs extract spatially local features from sequence windows, GCNs model topological relationships in protein structures or interaction networks, and transformers capture long-range dependencies through self-attention mechanisms. These architectural advances enable models to learn richer representations of protein function and context (Bileschi et al., 2022).

Three primary categories of ML-based function prediction are prominent in the literature (Figure **2**). First, EC number classifiers offer hierarchical predictions of enzymatic activity based on the chemical transformations enzymes catalyze. These models facilitate the classification of enzymes according to the four-level EC hierarchy, ranging from broad reaction class to specific substrate-product conversions (Kim et al., 2023). Second, GO term prediction models provide broader functional context, encompassing molecular functions, biological processes, and subcellular localization. These models often integrate structural, evolutionary, and ontological information (Kulmanov et al., 2024). Third, substrate specificity models predict enzyme-substrate compatibility, supporting rational enzyme design and high-throughput candidate prioritization. Such models can focus on binary classification (binds/does not bind), rank substrates by likelihood, or even predict complete substrate profiles (Gerloff et al., 2023; Yang et al., 2018).

Early EC classification models, such as ECPred (Dalkiran et al., 2018), relied on ensemble classifiers trained on curated sequence similarity data, demonstrating strong performance for well-represented enzyme classes. More recent models have adopted protein language model embeddings, including CLEAN (Yu et al., 2023b) and HiFi-NN (Ayres et al., 2023), which use ESM-derived representations to improve classification performance, especially for multifunctional or low-identity sequences. CLEAN employs contrastive learning to enhance embedding quality, while HiFi-NN uses hierarchical indexing for efficient similarity retrieval. Transformer-based architectures such as DeepECtransformer (Kim et al., 2023) and ECRECer (Shi et al., 2022) have further enhanced predictive capacity while introducing interpretability via attention mechanisms. DeepECtransformer addresses class imbalance using focal loss, and ECRECer combines recurrent units with attention in a multitask setup. Hybrid models like ProteInfer (Sanderson et al., 2023) and DeepFRI (Gligorijević et al., 2021) integrate CNNs and GCNs to simultaneously predict EC and GO terms, supporting multi-level functional annotations and enabling predictions for enzymes with little to no homology (see section S1 of Supplementary Material for more details).

For GO terms prediction, several models leverage deep learning and graph-based representations to capture the hierarchical nature of the Gene Ontology. DeepFRI applies GCNs to protein structure graphs, enabling spatial reasoning about functional sites (Gligorijević et al., 2021). PFresGO (Pan et al., 2023) incorporates the GO hierarchy into self-attention layers, while DeepGO-SE (Kulmanov et al., 2024) integrates symbolic reasoning by representing GO terms and protein sequences in a shared semantic space. Other models such as SPROF-GO (Yuan et al., 2023), TransFew (Boadu and Cheng, 2024), and PU-GO (Zhapa-Camacho et al., 2024) implement pre-trained embeddings from language models, few-shot learning to generalize from limited data, and positive-unlabeled frameworks to address label incompleteness (see section S1 of Supplementary Material for more details).

Substrate specificity models are increasingly central to enzyme mining, enabling the prioritization of candidate enzymes based on their predicted substrate profiles. From classical classifiers like GT-Predict (Yang et al., 2018) and AdenylPred (Robinson et al., 2020), to multimodal deep learning frameworks like ProSmith (Kroll et al., 2024) and ESP (Kroll et al., 2023), these models leverage both protein and substrate representations. ProSmith uses transformer-based embeddings on a multimodal transformer network and gradient-boosted decision trees to predict substrate profiles across thousands of enzyme-substrate pairs (Kroll et al., 2024). ESP integrates graph neural networks to encode substrate structures and combines them with enzyme embeddings for flexible specificity prediction via gradient boosting, too (Kroll et al., 2023). SEP-based models such as SEP-DNN (Watanabe et al., 2022) and EnzyPick (Xing et al., 2024) further extend predictive scope by incorporating product information, allowing full reaction outcome prediction. In parallel, hybrid ML pipelines for specific enzyme families, such as the model developed for bacterial nitrilases by Mou et al. (2020), demonstrate the utility of combining docking features, physicochemical descriptors, and ensemble learners for functional screening (see section S2 of Supplementary Material for more details).

ML-based annotation models provide scalable and accurate strategies for enzyme function prediction, serving as foundational tools in modern enzyme mining workflows (Ding et al., 2024). Their ability to operate across functional levels and data modalities facilitates the rapid screening and prioritization of candidate enzymes, accelerating the discovery of novel biocatalysts across sequence-diverse datasets (Landwehr et al., 2025).

### 3.2.2 Machine learning models for predicting enzymatic properties

Machine learning models have become increasingly central to the prediction of enzymatic properties that are critical for understanding enzyme function and optimizing their application in industrial contexts (Siedhoff et al., 2020). These include kinetic parameters, thermophilicity, pH optima, solubility, and other physicochemical characteristics (Figure **3**) (Li et al., 2022a). By providing scalable, high-throughput alternatives to traditional biochemical assays, ML-based approaches significantly enhance enzyme mining workflows, enabling the rapid identification and prioritization of candidate biocatalysts from large and diverse sequence datasets (Norton-Baker et al., 2024). These models can leverage diverse forms of input data, such as amino acid sequences, protein structural information, and environmental variables, and are implemented through a range of architectures, from conventional regression models to advanced neural networks and transformer-based systems (Boorla and Maranas, 2025; Yang et al., 2024a).

Among the most widely modeled enzymatic properties are kinetic constants such as $K_m$ and $k_{cat}$, which describe substrate affinity and catalytic turnover, respectively. Early predictive efforts employed regression algorithms using handcrafted features derived from sequences or molecular descriptors. For example, Mellor et al. (2016) employed Gaussian processes to estimate $K_m$ values based on reaction signatures. More advanced models such as DLKcat (Li et al., 2022a) integrated graph neural networks for substrate repre-sentation and CNN-based attention mechanisms for protein encoding. UniKP (Yu et al., 2023a) mitigated data imbalance through label distribution smoothing, while CatPred (Boorla and Maranas, 2025) incorporated 3D structural features to improve out-of-distribution performance. Multimodal frameworks like MPEK (Wang et al., 2024) leveraged pretrained language models including ProtT5 (Elnaggar et al., 2020) and Mole-BERT (Li and Jiang, 2021) to jointly predict $K_m$ and $k_{cat}$ across diverse enzyme classes (see section S3 of Supplementary Material for more details).

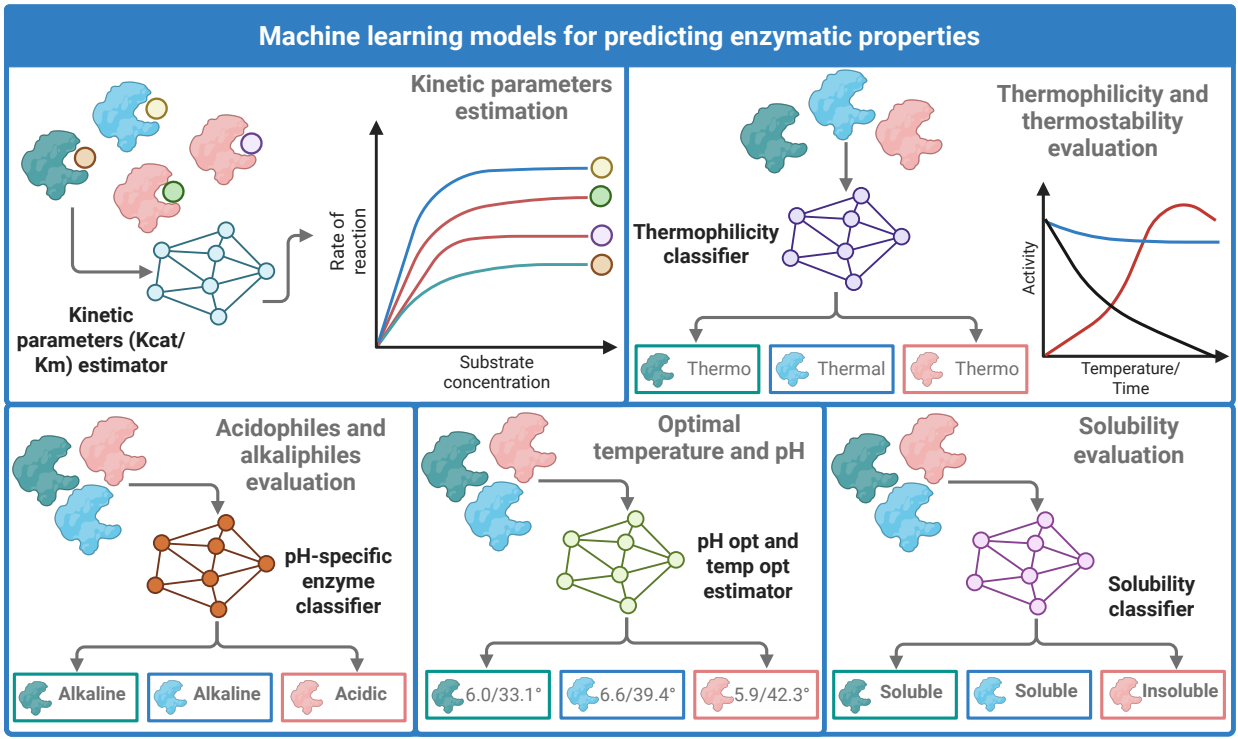

**Figure 3**: **Machine learning-based prediction of enzymatic physicochemical and kinetic properties**. (*Top left*) Estimation of kinetic constants such as $k_{cat}$ and $K_m$, enabling the prediction of catalytic efficiency from sequence and structural representations. (*Top right*) Thermostability and thermophilicity prediction based on physicochemical features and pre-trained embeddings, supporting the identification of enzymes suitable for elevated-temperature environments. (*Bottom left*) Classification of acidophilic and alkaliphilic enzymes using models trained to infer pH preference from amino acid composition and sequence motifs. (*Bottom center*) Prediction of optimal pH ($pH_{opt}$) and temperature ($T_{opt}$) conditions for enzymatic activity, allowing adaptation to specific industrial or environmental contexts. (*Bottom right*) Solubility classifiers predicting the likelihood of successful soluble expression in heterologous systems, aiding protein production pipelines. These models collectively support high-throughput enzyme mining by enabling the data-driven identification, prioritization, and characterization of functionally relevant biocatalysts.

Models like SEP-DNN (Watanabe et al., 2022) extended predictive capabilities by incorporating enzyme, substrate, and product information, moving beyond binary interaction predictions. These architectures enable more nuanced inferences, including potential product formation, thereby enhancing downstream utility in synthetic biology and biocatalysis.

Thermophilicity prediction has also benefited from ML advancements. Traditional classifiers such as SVMs and random forests demonstrated strong performance using feature engineering (Feng et al., 2020; Sharma et al., 2019; Foroozandeh Shahraki et al., 2021). Multi-class predictors like ThermoProt (Erickson et al., 2022) expanded the scope to include mesophilic and psychrophilic classes. Ensemble methods such as SAPPHIRE (Charoenkwan et al., 2022) further improved accuracy by integrating multiple learners.

Deep learning frameworks such as DeepTP (Zhao et al., 2023) and DeepPPThermo (Fang et al., 2019) introduced attention-based LSTM and CNN modules capable of capturing hierarchical sequence patterns. ProLaTherm (Haselbeck et al., 2023) demonstrated the effectiveness of contextual embeddings derived from ProtT5, outperforming models like BertThermo (Pei et al., 2023). ThermoFinder (Yu and Luo, 2024) pushed predictive accuracy beyond 98% by combining multiple pretrained embeddings through a meta-XGBoost ensemble (see section S4 and S5 of Supplementary Material for more details).

In the context of acidophilic and alkaliphilic enzyme classification, early models such as those by Zhang et al. (2009) and Su et al. (2015) used ensemble frameworks with structural and physicochemical features. Tools like AcalPred (Lin et al., 2013) and piSAAC (Khan et al., 2021) advanced this field with SVM and DNN-based predictors, achieving high classification accuracy and enabling rapid screening through accessible interfaces (see section S6 of Supplementary Material for more details).

Quantitative prediction of $T_{opt}$ and $pH_{opt}$ has become increasingly important. Support Vector Regressor (SVR) and Artificial Neural Networks (ANN) models tailored for specific enzyme classes (Chu et al., 2016; Yan and Wu, 2019) established early feasibility. TOME (Li et al., 2019) and TOMER (Gado et al., 2020) demonstrated the value of integrating organism-level data. Deep learning-based models such as Preoptem (Zhang et al., 2022b), EpHod (Gado et al., 2025) and Seq2Topt (Qiu et al., 2025) eliminated this dependency by using sequence embeddings and attention-based architectures. EpHod, in particular, offers interpretable residue-level insights that are valuable for protein engineering. Seq2Topt extents applicability by delivering a base architecture, well suited for adaption to other prediction task, showcased on optimal pH (Seq2pHopt) and melting temperature (Seq2Tm) prediction.

Dual-property predictors, such as MeTarEnz (Jahanshahi et al., 2023), integrate support vector regression and random forest algorithms to jointly estimate optimal temperature ($T_{opt}$) and pH ($pH_{opt}$). This approach has enabled the identification of thermally stable and alkaliphilic enzymes, including PersiLipase1, a promising candidate for high-temperature and alkaline industrial applications (see section S7 of Supplementary Material for more details).

Solubility prediction has also evolved from traditional models like Protein-Sol (Hebditch et al., 2017), SoDoPE (Bhandari et al., 2020), and SoluProt (Hon et al., 2021) to advanced deep learning architectures. DeepSol (Khurana et al., 2018), DSResSol (Madani et al., 2021), and DDcCNN (Wang et al., 2021) employed CNNs with residual and dilated connections to capture solubility-relevant features. Transformer-based models such as NetSolP (Thumuluri et al., 2021) and graph-based predictors like GraphSol (Chen et al., 2021), HybridGCN (Chen et al., 2023b), and PPSol (Chen et al., 2023a) demonstrated superior performance by integrating pretrained embeddings and structural contact maps (see section S8 of Supplementary Material for more details).

These ML models represent a transformative shift in how enzymatic properties are inferred from biological data. Their integration into enzyme mining platforms supports rapid, data-driven identification of functional enzymes tailored for specific industrial, environmental, and biomedical applications.

## 3.3 Current limitations in ML for enzyme mining

Despite rapid advances in ML for enzyme mining, significant challenges remain that limit the scalability, interpretability, and generalizability of predictive models (Medina-Ortiz et al., 2024c). One of the most critical and recurring challenges is experimental bias, resulting from the over representation of well-characterized enzymes in public datasets (Joshi et al., 2023). These imbalances biases toward learning algorithms toward dominant enzyme families and reduce model performance on understudied or novel proteins (Yu et al., 2023b).

In the context of functional classification, most models rely exclusively on sequence-based representations, limiting their ability to capture the complex multimodal nature of enzymatic function, which depends on the interplay of substrates, products, cofactors, and environmental factors (Hou et al., 2025). Only a handful of recent models incorporate substrate or product information (Kroll et al., 2023; Zhang et al., 2023; Heid et al., 2023; Probst et al., 2022), with ESMDance (Hou et al., 2025) standing out as one of the few efforts to explicitly learn physicochemical determinants of catalysis. Compounding this limitation is the lack of standardized benchmark datasets, which constrains the consistency and rigor of model evaluation and comparison (Notin et al., 2023). Many studies rely on datasets originally curated for unrelated tasks, such as variant effect prediction (Yang et al., 2024b; Xu et al., 2022; Groth et al., 2023; Dallago et al., 2021), resulting in limited coverage of the full enzymatic functional space.

For Gene Ontology prediction tasks, the hierarchical and interdependent nature of GO terms introduces further complexity. Capturing relationships among terms across molecular function, biological process, and cellular component domains requires advanced neural architectures and hierarchical loss formulations (Merino et al., 2022). In addition, computational demands for training and deploying these deep models, particularly with large-scale embeddings and attention mechanisms, remain nontrivial (Zhang et al., 2025).

The prediction of enzyme-substrate specificity introduces its own set of limitations. ML models still struggle to generalize to rare or previously unseen substrates (Campbell, 2024). Datasets are often biased toward industrially relevant enzyme substrate combinations, obscuring promiscuous or novel catalytic profiles (Nam et al., 2012). Inconsistent or incomplete annotations—especially in the form of missing negative examples or undocumented substrate ranges further compromise model reliability (Kroll et al., 2023; Gorantla et al., 2023).

For kinetic parameter prediction, such as $K_m$ and $k_{cat}$, several modeling bottlenecks persist. Enzyme kinetics are context-dependent and modulated by cofactors, temperature, pH, and other very enzyme specific experimental conditions, posing a challenge to generalize towards different enzyme classes (Wang et al., 2025b). Moreover, most ML models lack access to such metadata or fail to incorporate it systematically. Additionally, most kinetic data remains sparse and fragmented, often concentrated around well-studied enzyme classes (Yu et al., 2023a). High-dimensional embeddings, although powerful, can hinder interpretability, making it difficult to derive mechanistic insights from predictions (Medina-Ortiz et al., 2024c).

In the domain of thermostability and thermophilicity, training data is frequently skewed toward either ambient-temperature enzymes or extremophiles from polar and hydrothermal environments (Dou et al., 2023). This results in multimodal data distributions that complicate generalization. Moreover, models risk learning taxonomic or ecological proxies rather than the true biophysical determinants of thermal adaptation (Yang et al., 2024a).

Dataset bias and limited taxonomic representation likewise challenge the classification of acidophilic and alkaliphilic enzymes. Many classifiers are trained on a small number of enzyme families or microbial hosts, reducing their transferability to metagenomic datasets (Lin et al., 2013). Furthermore, models trained solely on annotated sequences often underperform on *de novo* sequences without close homologs, a common scenario across all predictors so far (Wang et al., 2025d).

Similar issues affect $T_{opt}$ and $pH_{opt}$ prediction, where early models were enzyme-specific and lacked generalizability. More recent models incorporating organismal metadata, such as optimal growth temperature (OGT), are limited by the availability of such data in non-model organisms (Li et al., 2019; Gado et al., 2020). Sequence-only models are broadly applicable but often exhibit reduced predictive precision (Zhang et al., 2022b). Even models with attention mechanisms, like EpHod (Gado et al., 2025), are constrained by the sparsity of training data for rare enzymes.

In solubility prediction, most existing models are trained on expression data from a limited number of hosts, especially *E. coli* (Hebditch et al., 2017; Hon et al., 2021). As a result, their generalizability to other systems remains uncertain. Tools like UniRep-RF (Martiny et al., 2021) represent a step toward host-specific modeling, but broader validation is needed. Additionally, differences between *in silico* predictions and *in vitro* outcomes underscore the necessity for more representative training datasets and experimental metadata. While deep models such as NetSolP (Thumuluri et al., 2021) and PPSol (Chen et al., 2023a) have improved accuracy, their high computational costs may pose constraints for routine deployment.

To address these limitations, future work should prioritize: 1) the curation of larger, more diverse, and balanced datasets across enzyme classes and conditions; 2) the integration of multi-modal data sources, including structure, expression system, and environmental metadata; and 3) the development of interpretable, efficient, and transferable ML architectures capable of operating in low-data regimes. Hybrid modeling approaches that combine deep embeddings with classical ML classifiers (Liaqat et al., 2021), and attention-based interpretability frameworks (Li et al., 2022b), offer promising directions. As these challenges are incrementally addressed, ML models will become increasingly indispensable for the scalable, accurate, and interpretable discovery and characterization of enzymes.

### 3.4 Data-driven frameworks for enzyme mining: Demonstrative case examples

By enabling the prioritization and discovery of enzymes with specific catalytic functions, ML-based strategies address the inherent limitations of conventional screening methods—such as reliance on cultivable organisms, low throughput, and annotation bias. This section highlights selected case studies that exemplify the successful integration of ML models into enzyme mining pipelines, focusing on applications related to plastic degradation, mycotoxin detoxification, terpene biosynthesis, and phage lysin identification. These examples collectively underscore the power of ML to rapidly interrogate vast sequence repositories, uncover novel enzymatic functions, and guide experimental validation.

In the context of plastic degradation, Danso et al. (2018) developed an SVM-based model to identify PET hydrolases, trained on a curated dataset of experimentally validated sequences. The model demonstrated robust predictive performance, facilitating the detection of candidate enzymes with potential PET-degrading activity. Building upon this foundation, Zhang et al. (2022a) employed a Hidden Markov Model (HMM) approach that incorporated domain-specific motifs to improve specificity and functional annotation, further enhancing the accuracy of PET hydrolase prediction. More recently, Medina-Ortiz et al. (2025) introduced an AI-driven framework that combines protein language models with generative design to accelerate the discovery and engineering of plastic-degrading enzymes. By fine-tuning pre-trained models for PET specificity, the authors identified over 6,000 candidate hydrolases—many active *in silico*—with average classification accuracy exceeding 89%, demonstrating the cutting-edge role of ML in expanding PET biocatalyst discovery.

Expanding to broader classes of plastic-degrading enzymes, Jiang et al. (2023) implemented a XGBoost-based classifier to identify enzymes capable of degrading various polymers, including polyethylene and polystyrene. The model achieved high predictive accuracy and was instrumental in nominating candidates for experimental testing. In a follow-up study, Jin and Jia (2024) applied the framework to identify a novel esterase with confirmed polystyrene microplastic-degrading activity, thus validating the practical utility of ML-guided enzyme discovery.

For mycotoxin-degrading enzymes, the PU-EPP model introduced by Zhang et al. (2023) employed a positive-unlabeled (PU) learning framework to predict enzyme–substrate interactions, specifically targeting ochratoxin A (OTA) and zearalenone (ZEA). The model, trained on a dataset comprising over 170,000 enzymes and more than 600,000 enzyme–substrate pairs, addressed class imbalance through a weighted sampling strategy. Remarkably, 15 of the top 20 predicted enzymes were experimentally confirmed to degrade OTA and ZEA, with six achieving over 90% degradation within three hours—demonstrating the model's precision and functional relevance.

The discovery of terpene synthases (TPSs) by Samusevich et al. (2024) illustrates the synergy between protein language models, structural prediction, and ML classification. Leveraging a Random Forest classifier trained on curated TPS sequences and embedding representations, the model achieved high precision in both TPS identification and substrate specificity prediction. Experimental validation using the UniRef50 database yielded 17 novel TPS candidates, seven of which were confirmed to exhibit enzymatic activity.

In the domain of phage lysins, Fu et al. (2024) developed DeepMineLys, a convolutional neural network (CNN) model that combines dual-track embeddings capturing both physicochemical and sequence-derived features. Applied to human microbiome datasets, the model achieved an F1-score of 84% on an independent validation set. From 624 predicted non-redundant lysin candidates, 11 were confirmed to be enzymatically active, including one variant with 6.2-fold greater activity than hen egg white lysozyme.

These case studies illustrate the growing role of machine learning in transforming enzyme mining from a labor-intensive and low-throughput process into a data-driven, predictive, and scalable framework. The integration of diverse datasets, sophisticated model architectures, and rigorous experimental validation enables the identification of functionally novel enzymes across broad sequence spaces. As these methodologies continue to evolve—particularly through the incorporation of multi-modal learning, transfer learning, and interpretable AI—the landscape of enzyme discovery is poised to expand rapidly, offering powerful tools for addressing industrial, pharmaceutical, and environmental challenges through biocatalysis.

# 4 Towards autonomous enzyme discovery through ML-guided mining platforms

This section introduces a modular, ML-guided strategy for enzyme mining that integrates representation learning, functional classification, and property estimation to enable autonomous candidate discovery. We outline the proposed pipeline and examine key challenges—such as data quality, interpretability, and generalizability—that must be addressed to realize fully automated enzyme mining workflows.

## 4.1 A machine learning-guided framework for enzyme mining

To address the limitations of traditional enzyme mining workflows and overcome the challenges identified in ML-based prediction strategies, we propose an integrated and modular framework that leverages advances in machine learning to enable scalable, data-driven discovery of functional enzymes from genomic and metagenomic datasets (Figure **4**). This approach aims not only to improve prediction accuracy and generalizability but also to enhance candidate prioritization, facilitate experimental design, and provide an integrated interface with enzyme engineering pipelines.

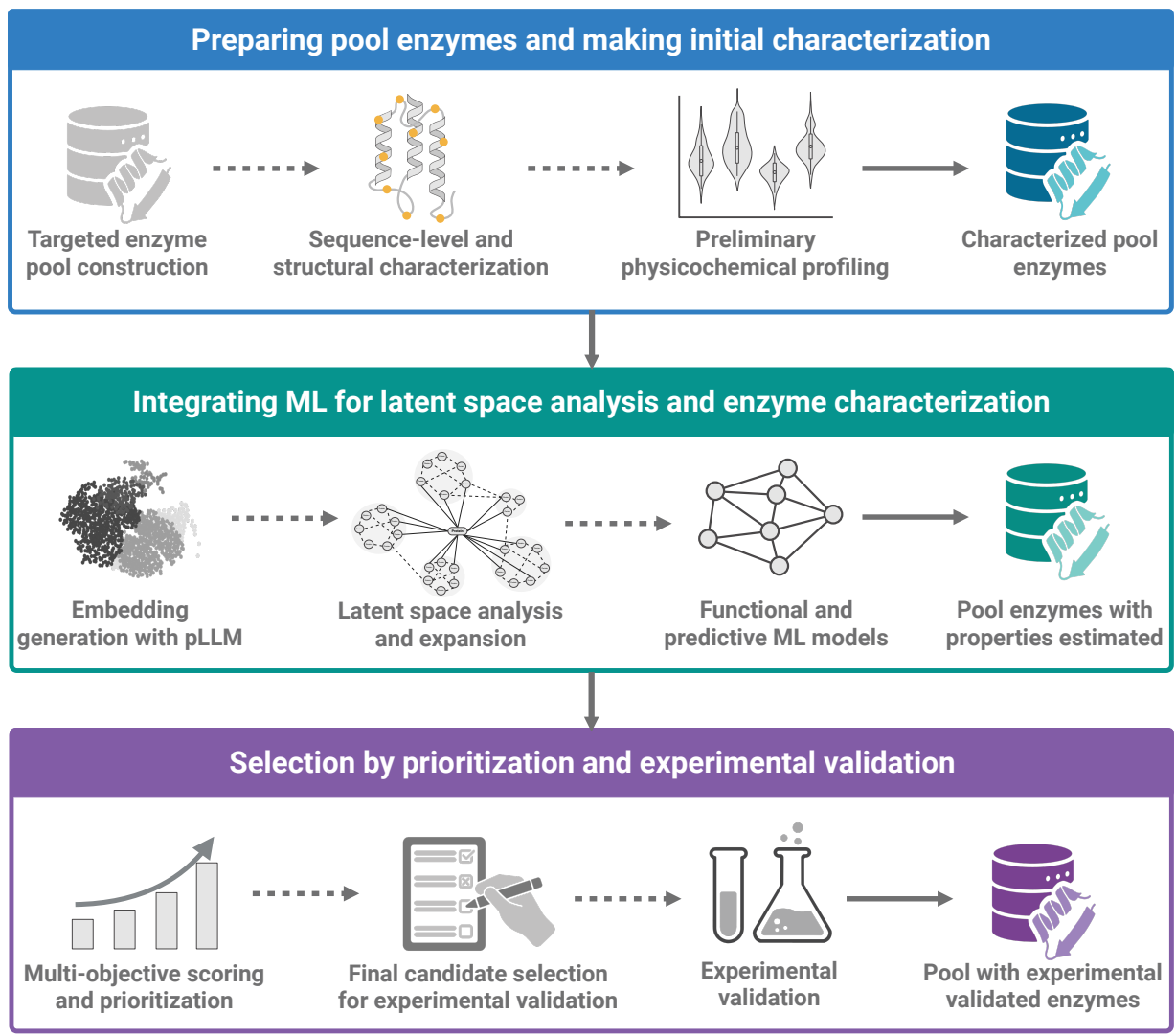

**Figure 4**: **Machine learning-guided framework for autonomous enzyme mining and prioritization.** The process begins with the construction of a targeted enzyme pool, assembled from genomic and metagenomic databases using structural, functional, and sequence-based queries. Sequence-level and structural characterization—through clustering, phylogenetic mapping, or sequence similarity networks—enables visualization of diversity and identification of candidate clusters. Pretrained protein language models are used to generate embeddings, which are projected into a latent space for the detection of underexplored or functionally divergent regions. Latent space clustering supports candidate expansion and guides the application of predictive models for functional annotation (e.g., EC number, GO terms, substrate specificity) and property estimation (e.g., $K_m$, $k_{cat}$, $T_{opt}$, $pH_{opt}$, solubility). These predictions inform a multi-objective ranking system that prioritizes enzymes based on novelty, promiscuity, and desired traits. Top candidates are selected for experimental validation, and empirical results are reintegrated to refine model performance, establishing a closed-loop discovery framework optimized for scalability and precision.

The framework begins with the generation of a tailored enzyme pool, constructed by querying protein databases or annotated genomic and metagenomic datasets according to structural, functional, or sequence-based criteria. Leveraging genome mining tools and domain-specific filters, this step enables the extraction of a comprehensive set of enzyme sequences aligned with a given biocatalytic goal. When applied to metagenomes, this strategy allows access to uncultured or extremophilic organisms, offering a vast and untapped source of enzymatic diversity (Foroozandeh Shahraki et al., 2021). Sequence retrieval can be augmented with homology-based searches, domain architectures, or the use of AI-curated repositories enriched by structural predictions or metaproteomic evidence (Yeo et al., 2025).

Subsequent characterization of the enzyme pool employs sequence similarity networks (SSNs), phylogenetic trees, or graph-based approaches to visualize diversity, identify evolutionary clusters, and reduce redundancy (Oberg et al., 2023; Atkinson et al., 2009; Fuchs and Höpken, 2022). These representations allow researchers to evaluate the coverage of known enzymatic space and detect potential functional outliers or unexplored clusters.

To enhance the discovery of non-trivial candidates, the enzyme pool is projected into a latent space defined by embeddings derived from pre-trained protein language models (Rives et al., 2021). This dimensionality reduction captures complex sequence-function relationships, enabling the detection of patterns that may be missed by classical alignments. Latent space analysis facilitates candidate expansion through similarity searches or clustering techniques, revealing enzymes located in poorly explored or functionally ambiguous regions of the embedding space (Bileschi et al., 2022).

Clustering in latent space plays a pivotal role in identifying sequence groups that are either underrepresented in training data or distant from known functional classes. This stage supports the discovery of enzymes that may possess novel or promiscuous activities, serving as reservoirs of biocatalytic innovation (Xu et al., 2022).

Before or after expanding and clustering, the enzyme pool is subjected to a suite of functional classification models, including EC number, GO term, and substrate specificity prediction. These classifiers, built upon diverse ML architectures such as transformers, graph neural networks, and hybrid models, provide hierarchical and multifaceted annotations, enabling rapid screening of candidate functions across biochemical domains (Sanderson et al., 2023; Kulmanov et al., 2024). In parallel, ML-based estimators are used to predict enzymatic properties such as optimal temperature, pH, thermostability, solubility, and kinetic parameters (e.g., $K_m$, $k_{cat}$) (Boorla and Maranas, 2025).

The outputs from functional classifiers and property estimators are integrated to construct a candidate ranking system, guided by multi-objective scoring and prioritization approaches. This system prioritizes enzymes based on novelty (distance from known clusters), predicted promiscuity (multi-substrate compatibility or multifunctional potential), and alignment with user-defined application criteria (e.g., thermostability thresholds, substrate preferences). By balancing these multiple objectives, the framework enables the systematic identification of candidates with both functional relevance and engineering potential. This stage also facilitates the recognition of enzyme families that may benefit from directed evolution or rational design strategies (Ariaeenejad et al., 2024).

Final candidate selection is informed by a combination of latent space coverage and predicted performance. By optimizing diversity and predicted function, the selected enzyme panel maximizes the likelihood of experimental success. These candidates are subjected to *in vitro* validation, including expression, purification, and biochemical assays. Experimental data, when available, are reintegrated into the pipeline to refine model performance and guide future iterations, establishing a closed feedback loop that enhances robustness and adaptability (Hecko et al., 2023).

This ML-guided strategy is particularly well suited for integration with enzyme engineering workflows. Candidates discovered through this pipeline may exhibit suboptimal traits for direct application but possess latent potential that can be harnessed through rational or data-driven optimization (Zhou et al., 2024). For example, enzymes predicted to catalyze desired reactions but lacking thermal resilience may be targeted for thermostabilization through mutagenesis or structure-based redesign. In this way, enzyme mining and engineering are connected as complementary processes within a unified, ML-enhanced discovery architecture.

## 4.2 Opportunities and remaining challenges

The integration of ML into enzyme mining pipelines marks a significant advance in the discovery and characterization of biocatalysts with tailored properties (Markus et al., 2023). While traditional approaches grounded in sequence homology and manual curation have yielded valuable insights, they are inherently constrained by their limited resolution across distant evolutionary relationships and the difficulty of scaling to increasingly vast genomic and metagenomic datasets (Zhou et al., 2024). In contrast, ML enables the extraction of functional signatures from complex data spaces, offering predictive frameworks capable of inferring enzyme activity, specificity, and physicochemical traits (Landwehr et al., 2025). Yet, the path toward fully autonomous, ML-guided enzyme mining systems presents both considerable opportunities and critical technical challenges.

One of the most promising opportunities lies in the ability of ML models to prioritize candidate sequences based on likelihood of desired biochemical attributes—including substrate scope, catalytic turnover, thermal stability, and pH optima—at unprecedented scales (Siedhoff et al., 2020). By integrating multi-modal representations derived from sequences, structures, and molecular descriptors, these models facilitate targeted selection of enzymes for industrial, pharmaceutical, and environmental applications (Boorla and Maranas, 2025). Recent examples, such as DeepMineLys and SEP-DNN, exemplify the effectiveness of ML frameworks in identifying novel enzymes with experimentally validated functionality (Fu et al., 2024; Watanabe et al., 2022). Similarly, tools like GT-Predict and AdenylPred demonstrate how the inclusion of substrate-centric features enhances the functional resolution of predictions, enabling the prioritization of candidates with broader or more selective substrate ranges (Yang et al., 2018; Robinson et al., 2020). Extending such strategies to undercharacterized enzyme families and microbial taxa remains a critical avenue for expanding the functional enzyme space.

However, these advancements are limited by persistent challenges, most notably the limitations of available training data. Many ML models are trained on reduced subsets of enzymes derived from extensively studied organisms, resulting in dataset bias and reduced generalizability to novel or low-homology sequences (Ayres et al., 2023). This issue is particularly acute in substrate-specificity modeling, where the scarcity of reliable negative data complicates classifier calibration (Kroll et al., 2023). While positive-unlabeled learning strategies, such as those employed by PU-EPP, offer a promising mitigation pathway, methodological refinements are needed to prevent model overfitting and information leakage (Cai et al., 2025). In parallel, the reliance on curated and often redundant databases risks perpetuating knowledge gaps, inadvertently reinforcing focus on well-established enzyme functions while neglecting mechanistically novel or ecologically relevant activities (Nikoladatos et al., 2022).

Model interpretability presents a further challenge. Although transformer-based and graph-based architectures have markedly improved predictive performance across various tasks, the high-dimensional embeddings they generate often lack direct biological transparency (Medina-Ortiz et al., 2024c; García-Vinuesa et al., 2025). Some advances—such as the use of attention mechanisms in DeepECtransformer or PFresGO—help localize salient residues or motifs associated with specific functions (Kim et al., 2023; Pan et al., 2023). Nevertheless, a systematic framework for translating these insights into mechanistic understanding remains underdeveloped, limiting the utility of such models for hypothesis generation or rational enzyme design (Hunklinger and Ferruz, 2025).

Several strategies may help address these limitations and advance the development of robust, autonomous ML-assisted enzyme mining workflows. Expanding and diversifying training datasets—especially through the functional annotation of metagenomic data—will be essential for improving model generalizability and minimizing taxonomic and functional biases (Yeo et al., 2025). Multi-task learning architectures that jointly predict multiple enzymatic traits (e.g. function, specificity, kinetics, and solubility) offer a path toward more coherent and efficient modeling pipelines (Wang et al., 2024, 2025a). In parallel, hybrid model architectures that integrate interpretable machine learning methods with transformer-derived embeddings may help strike a balance between performance and transparency (Hunklinger and Ferruz, 2025).

Moreover, the strategic incorporation of genome and metagenome data into this framework offers a unique opportunity to mine functional diversity from previously untapped ecological niches (Robinson et al., 2021). Integrating ML predictions with phylogenetic, structural, and environmental metadata can support the prioritization of enzyme candidates that are not only functionally relevant but also amenable to downstream optimization (Yeo et al., 2025). In this context, the proposed pipeline may serve as a precursor to enzyme engineering efforts, selecting candidate sequences that are primed for rational modification or experimental evolution.

Finally, the convergence of deep learning, data-centric workflows, and high-throughput experimental techniques is poised to redefine the landscape of enzyme mining. As predictive models mature and interpretability tools evolve, the field moves closer to autonomous platforms capable of guiding biocatalyst discovery across vast and uncharted regions of protein space (Hon et al., 2020). Addressing the current challenges—particularly those related to data quality, model transparency, and benchmarking—will be critical for realizing the full potential of ML-driven enzyme mining in delivering functionally diverse and industrially robust enzymes.

# 5   Outlook and future perspectives

The incorporation of machine learning into enzyme mining workflows marks a pivotal advancement in the discovery and functional characterization of biocatalysts across diverse domains. This work has provided a comprehensive examination of ML-based models designed to infer enzymatic function and properties, including tools for EC number classification, Gene Ontology term assignment, substrate specificity prediction, and physicochemical property estimation such as kinetic parameters, solubility, thermophilicity, and pH optima. In addition to outlining the architectural diversity of these models—from classical algorithms to deep neural networks and protein language models—we have showcased representative case studies demonstrating the practical utility of ML frameworks in accelerating enzyme discovery for biotechnological applications.

To address the aforementioned challenges, future efforts should emphasize the development of multi-task learning frameworks capable of jointly predicting functional annotations and enzymatic properties that are of interest. Expanding and diversifying training datasets—particularly through the integration of metagenomic resources—and employing strategies such as positive-unlabeled learning, data augmentation, and transfer learning will be essential for improving robustness and coverage. At the same time, embedding explainability into model design through attention mechanisms, saliency mapping, or symbolic reasoning will enhance the transparency and interpretability of predictions, facilitating experimental validation and broader adoption.

As the field moves toward autonomous enzyme discovery platforms, the convergence of ML models, high-throughput experimental workflows, and modular decision-making systems offers a scalable pathway to uncover functionally novel enzymes with tailored properties. Establishing standardized benchmarks, interoperable pipelines, and shared community resources will be critical to ensuring reproducibility, comparability, and sustained innovation. Collectively, these developments position ML-guided enzyme mining not only as a complement to traditional discovery methods, but as a foundational paradigm for next-generation biocatalyst development across industrial, environmental, and biomedical settings.

## Conflict of interest statement

The authors declare that the research was conducted without any commercial or financial relationships that could be construed as a potential conflict of interest.

## Data availability statement

This work does not produce, collect or use datasets, training models, and novel results.

## Author contributions statement

MDD, JZ, DM-O: conceptualization. YZ, FM, SQ, WD: investigation. YZ, FM, DM-O, SQ, WD: writing. All authors: review and editing. MDD, JZ: supervision and funding resources. MDD, JZ: project administration. All authors have read and agreed to the published version of the manuscript.

## Acknowledgments

This work was supported by the National Key Research and Development Program of China (No. 2019YFA0906400), Science and Technology Innovation Talent Program of Hubei Province (No. 2023DJC122). MDD acknowledges funding by the Deutsche Forschungsgemeinschaft (DFG, German Research Foundation) - within the Priority Program 840 Molecular Machine Learning SPP2363 (Project Number 497207454) and the Bundesministerium für Bildung und Forschung (BMBF) project (FKZ: 031B1442A).

## References

Ahmed, N., Zhang, B., Deng, L., Bozdar, B., Li, J., Chachar, S., Chachar, Z., Jahan, I., Talpur, A., Gishkori, M. S., et al. (2024). Advancing horizons in vegetable cultivation: a journey from ageold practices to high-tech greenhouse cultivation—a review. *Frontiers in Plant Science*, 15:1357153.

Albright, S. and Louca, S. (2023). Trait biases in microbial reference genomes. *Scientific Data*, 10(1):84.

Alla, S. and Adari, S. K. (2020). What is mlops? In *Beginning MLOps with MLFlow: Deploy Models in AWS SageMaker, Google Cloud, and Microsoft Azure*, pages 79–124. Springer.

Altschul, S. F., Madden, T. L., Schäffer, A. A., Zhang, J., Zhang, Z., Miller, W., and Lipman, D. J. (1997). Gapped blast and psi-blast: a new generation of protein database search programs. *Nucleic acids research*, 25(17):3389–3402.

Andersen, L. K. and Reading, B. J. (2024). A supervised machine learning workflow for the reduction of highly dimensional biological data. *Artificial Intelligence in the Life Sciences*, 5:100090.

Ao, Y.-F., Dörr, M., Menke, M. J., Born, S., Heuson, E., and Bornscheuer, U. T. (2024). Data-driven protein engineering for improving catalytic activity and selectivity. *ChemBioChem*, 25(3):e202300754.

Ariaeenejad, S., Gharechahi, J., Foroozandeh Shahraki, M., Fallah Atanaki, F., Han, J.-L., Ding, X.-Z., Hildebrand, F., Bahram, M., Kavousi, K., and Hosseini Salekdeh, G. (2024). Precision enzyme discovery through targeted mining of metagenomic data. *Natural Products and Bioprospecting*, 14(1):7.

Atkinson, H. J., Morris, J. H., Ferrin, T. E., and Babbitt, P. C. (2009). Using sequence similarity networks for visualization of relationships across diverse protein superfamilies. *PLOS ONE*, 4(2):1–14.

Ayres, G., Munsamy, G., Heinzinger, M., Ferruz, N., Yang, K., and Lorenz, P. (2023). Hifi-nn annotates the microbial dark matter with enzyme commission numbers. In *Machine Learning for Structural Biology Workshop, NeurIPS*.

Bauman, K. D., Butler, K. S., Moore, B. S., and Chekan, J. R. (2021). Genome mining methods to discover bioactive natural products. *Natural product reports*, 38(11):2100–2129.

Bhandari, B. K., Gardner, P. P., and Lim, C. S. (2020). Solubility-weighted index: fast and accurate prediction of protein solubility. *Bioinformatics*, 36(18):4691–4698.

Bileschi, M. L., Belanger, D., Bryant, D. H., Sanderson, T., Carter, B., Sculley, D., Bateman, A., DePristo, M. A., and Colwell, L. J. (2022). Using deep learning to annotate the protein universe. *Nature Biotechnology*, 40(6):932–937.

Biswas, S., Khimulya, G., Alley, E. C., Esvelt, K. M., and Church, G. M. (2021). Low-n protein engineering with data-efficient deep learning. *Nature methods*, 18(4):389–396.

Boadu, F. and Cheng, J. (2024). Improving protein function prediction by learning and integrating representations of protein sequences and function labels. *Bioinformatics Advances*, 4(1).

Boger, R. S., Chithrananda, S., Angelopoulos, A. N., Yoon, P. H., Jordan, M. I., and Doudna, J. A. (2025). Functional protein mining with conformal guarantees. *Nature Communications*, 16(1):85.

Boorla, V. S. and Maranas, C. D. (2025). Catpred: a comprehensive framework for deep learning in vitro enzyme kinetic parameters. *Nature Communications*, 16(1).

Cai, P., Liu, D., Xing, H., Zhang, D., Le, Y., Wu, A., and Hu, Q.-N. (2025). Deepmbenzy: An ai-driven database of mycotoxin biotransformation enzymes. *Journal of Agricultural and Food Chemistry*.

Campbell, M. J. (2024). Viper: A general model for prediction of enzyme substrates. *bioRxiv*, pages 2024–06.

Charoenkwan, P., Schaduangrat, N., Moni, M. A., Lio', P., Manavalan, B., and Shoombuatong, W. (2022). Sapphire: A stacking-based ensemble learning framework for accurate prediction of thermophilic proteins. *Computers in Biology and Medicine*, 146:105704.

Chen, J., Qian, Y., Huang, Z., Xiao, X., and Deng, L. (2023a). Enhancing protein solubility prediction through pre-trained language models and graph convolutional neural networks. In *2023 IEEE International Conference on Bioinformatics and Biomedicine (BIBM)*. IEEE.

Chen, J., Zheng, S., Zhao, H., and Yang, Y. (2021). Structure-aware protein solubility prediction from sequence through graph convolutional network and predicted contact map. *Journal of Cheminformatics*, 13(1).

Chen, L., Wu, R., Zhou, F., Zhang, H., and Liu, J. K. (2023b). Hybridgcn for protein solubility prediction with adaptive weighting of multiple features. *Journal of Cheminformatics*, 15(1).

Chu, Y., Yi, Z., Zeng, R., and Zhang, G. (2016). Predicting the optimum temperature of $\beta$-agarase based on the relative solvent accessibility of amino acids. *Journal of Molecular Catalysis B: Enzymatic*, 129:47–53.

Dalkiran, A., Rifaioglu, A. S., Martin, M. J., Cetin-Atalay, R., Atalay, V., and Doğan, T. (2018). Ecpred: a tool for the prediction of the enzymatic functions of protein sequences based on the ec nomenclature. *BMC Bioinformatics*, 19(1).

Dallago, C., Mou, J., Johnston, K. E., Wittmann, B. J., Bhattacharya, N., Goldman, S., Madani, A., and Yang, K. K. (2021). Flip: Benchmark tasks in fitness landscape inference for proteins. *bioRxiv*, pages 2021–11.

Danso, D., Schmeisser, C., Chow, J., Zimmermann, W., Wei, R., Leggewie, C., Li, X., Hazen, T., and Streit, W. R. (2018). New insights into the function and global distribution of polyethylene terephthalate (pet)-degrading bacteria and enzymes in marine and terrestrial metagenomes. *Applied and Environmental Microbiology*, 84(8).

Daud, M., Charton, P., Damour, C., Wang, J., and Cadet, F. (2025). Ehcube4p: Learning epistatic patterns through hypercube graph convolution neural network for protein fitness function estimation. *arXiv preprint arXiv:2506.16921*.

David, K. T. and Halanych, K. M. (2023). Unsupervised deep learning can identify protein functional groups from unaligned sequences. *Genome Biology and Evolution*, 15(5):evad084.

Ding, K., Luo, J., and Luo, Y. (2024). Leveraging conformal prediction to annotate enzyme function space with limited false positives. *PLOS Computational Biology*, 20(5):e1012135.

Dou, Z., Sun, Y., Jiang, X., Wu, X., Li, Y., Gong, B., and Wang, L. (2023). Data-driven strategies for the computational design of enzyme thermal stability: Trends, perspectives, and prospects: Data-driven strategies for enzyme thermostability design. *Acta Biochimica et Biophysica Sinica*, 55(3):343.

Elnaggar, A., Heinzinger, M., Dallago, C., Rihawi, G., Wang, Y., Jones, L., Gibbs, T., Feher, T., Angerer, C., Steinegger, M., Bhowmik, D., and Rost, B. (2020). Prottrans: Towards cracking the language of life's code through self-supervised deep learning and high performance computing.

Erden, C., Demir, H. I., and Kökçam, A. H. (2023). Enhancing machine learning model performance with hyper parameter optimization: A comparative study. *arXiv preprint arXiv:2302.11406*.

Erickson, E., Gado, J. E., Avilán, L., Bratti, F., Brizendine, R. K., Cox, P. A., Gill, R., Graham, R., Kim, D.-J., König, G., Michener, W. E., Poudel, S., Ramirez, K. J., Shakespeare, T. J., Zahn, M., Boyd, E. S., Payne, C. M., DuBois, J. L., Pickford, A. R., Beckham, G. T., and McGeehan, J. E. (2022). Sourcing thermotolerant poly(ethylene terephthalate) hydrolase scaffolds from natural diversity. *Nature Communications*, 13(1).

Fang, X., Huang, J., Zhang, R., Wang, F., Zhang, Q., Li, G., Yan, J., Zhang, H., Yan, Y., and Xu, L. (2019). Convolution neural network-based prediction of protein thermostability. *Journal of Chemical Information and Modeling*, 59(11):4833–4843.

Feehan, R., Montezano, D., and Slusky, J. S. (2021). Machine learning for enzyme engineering, selection and design. *Protein Engineering, Design and Selection*, 34:gzab019.

Feng, C., Ma, Z., Yang, D., Li, X., Zhang, J., and Li, Y. (2020). A method for prediction of thermophilic protein based on reduced amino acids and mixed features. *Frontiers in Bioengineering and Biotechnology*, 8.

Foroozandeh Shahraki, M., Farhadyar, K., Kavousi, K., Azarabad, M. H., Boroomand, A., Ariaeenejad, S., and Hosseini Salekdeh, G. (2021). A generalized machine-learning aided method for targeted identification of industrial enzymes from metagenome: A xylanase temperature dependence case study. *Biotechnology and Bioengineering*, 118(2):759–769.

Fu, L., Niu, B., Zhu, Z., Wu, S., and Li, W. (2012). Cd-hit: accelerated for clustering the next-generation sequencing data. *Bioinformatics*, 28(23):3150–3152.

Fu, Y., Yu, S., Li, J., Lao, Z., Yang, X., and Lin, Z. (2024). Deepminelys: Deep mining of phage lysins from human microbiome. *Cell Reports*, 43(8):114583.

Fuchs, M. and Höpken, W. (2022). Clustering: Hierarchical, k-means, dbscan. In *Applied Data Science in Tourism: Interdisciplinary Approaches, Methodologies, and Applications*, pages 129–149. Springer.

Funk, J., Machado, L., Bradley, S. A., Napiorkowska, M., Gallegos-Dextre, R., Pashkova, L., Madsen, N. G., Webel, H., Phaneuf, P. V., Jenkins, T. P., et al. (2024). Proteusai: An open-source and user-friendly platform for machine learning-guided protein design and engineering. *bioRxiv*, pages 2024–10.

Gado, J. E., Beckham, G. T., and Payne, C. M. (2020). Improving enzyme optimum temperature prediction with resampling strategies and ensemble learning. *Journal of Chemical Information and Modeling*, 60(8):4098–4107.

Gado, J. E., Knotts, M., Shaw, A. Y., Marks, D., Gauthier, N. P., Sander, C., and Beckham, G. T. (2025). Machine learning prediction of enzyme optimum ph. *Nature Machine Intelligence*.

Gantz, M., Neun, S., Medcalf, E. J., van Vliet, L. D., and Hollfelder, F. (2023). Ultrahigh-throughput enzyme engineering and discovery in in vitro compartments. *Chemical Reviews*, 123(9):5571–5611.

García-Vinuesa, J., Rojas, J., Soto-García, N., Martínez, N., Alvarez-Saravia, D., Uribe-Paredes, R., Davari, M. D., Conca, C., Asenjo, J. A., and Medina-Ortiz, D. (2025). Geometric deep learning assists protein engineering. opportunities and challenges. *arXiv preprint arXiv:2506.16091*.

Gerloff, D. L., Ilina, E. I., Cialini, C., Salcedo, U. M., Mittelbronn, M., and Müller, T. (2023). Prediction and verification of glycosyltransferase activity by bioinformatics analysis and protein engineering. *STAR protocols*, 4(1):101905.

Gligorijević, V., Renfrew, P. D., Kosciolek, T., Leman, J. K., Berenberg, D., Vatanen, T., Chandler, C., Taylor, B. C., Fisk, I. M., Vlamakis, H., Xavier, R. J., Knight, R., Cho, K., and Bonneau, R. (2021). Structure-based protein function prediction using graph convolutional networks. *Nature Communications*, 12(1).

Goles, M., Daza, A., Cabas-Mora, G., Sarmiento-Varón, L., Sepúlveda-Yañez, J., Anvari-Kazemabad, H., Davari, M. D., Uribe-Paredes, R., Olivera-Nappa, Á., Navarrete, M. A., et al. (2024). Peptide-based drug discovery through artificial intelligence: towards an autonomous design of therapeutic peptides. *Briefings in Bioinformatics*, 25(4):bbae275.

Gorantla, R., Kubincová, A., Weiße, A. Y., and Mey, A. S. (2023). From proteins to ligands: decoding deep learning methods for binding affinity prediction. *Journal of Chemical Information and Modeling*, 64(7):2496–2507.

Groth, P. M., Michael, R., Salomon, J., Tian, P., and Boomsma, W. (2023). FLOP: Tasks for fitness landscapes of protein wildtypes. *bioRxiv*.

Han, S.-R., Park, M., Kosaraju, S., Lee, J., Lee, H., Lee, J. H., Oh, T.-J., and Kang, M. (2024). Evidential deep learning for trustworthy prediction of enzyme commission number. *Briefings in Bioinformatics*, 25(1):bbad401.

Harding-Larsen, D., Funk, J., Madsen, N. G., Gharabli, H., Acevedo-Rocha, C. G., Mazurenko, S., and Welner, D. H. (2024). Protein representations: Encoding biological information for machine learning in biocatalysis. *Biotechnology Advances*, 77:108459.

Haselbeck, F., John, M., Zhang, Y., Pirnay, J., Fuenzalida-Werner, J., Costa, R., and Grimm, D. (2023). Superior protein thermophilicity prediction with protein language model embeddings. *NAR Genomics and Bioinformatics*, 5(4).

Hebditch, M., Carballo-Amador, M. A., Charonis, S., Curtis, R., and Warwicker, J. (2017). Protein–sol: a web tool for predicting protein solubility from sequence. *Bioinformatics*, 33(19):3098–3100.

Hecko, S., Schiefer, A., Badenhorst, C. P., Fink, M. J., Mihovilovic, M. D., Bornscheuer, U. T., and Rudroff, F. (2023). Enlightening the path to protein engineering: chemoselective turn-on probes for high-throughput screening of enzymatic activity. *Chemical Reviews*, 123(6):2832–2901.

Heid, E., Probst, D., Green, W. H., and Madsen, G. K. H. (2023). EnzymeMap: Curation, validation and data-driven prediction of enzymatic reactions. *Chemical Science*.

Hon, J., Borko, S., Stourac, J., Prokop, Z., Zendulka, J., Bednar, D., Martinek, T., and Damborsky, J. (2020). Enzymeminer: automated mining of soluble enzymes with diverse structures, catalytic properties and stabilities. *Nucleic acids research*, 48(W1):W104–W109.

Hon, J., Marusiak, M., Martinek, T., Kunka, A., Zendulka, J., Bednar, D., and Damborsky, J. (2021). Soluprot: prediction of soluble protein expression in escherichia coli. *Bioinformatics*, 37(1):23–28.

Hou, C., Zhao, H., and Shen, Y. (2025). Learning biophysical dynamics with protein language models. *bioRxiv*, pages 2024–10.

Hunklinger, A. and Ferruz, N. (2025). Toward the explainability of protein language models for sequence design. *arXiv preprint arXiv:2506.19532*.

Jahanshahi, D. A., Ariaeenejad, S., and Kavousi, K. (2023). A metagenomic catalog for exploring the plastizymes landscape covering taxa, genes, and proteins. *Scientific Reports*, 13(1).

Jiang, R., Shang, L., Wang, R., Wang, D., and Wei, N. (2023). Machine learning based prediction of enzymatic degradation of plastics using encoded protein sequence and effective feature representation. *Environmental Science & Technology Letters*, 10(7):557–564.

Jin, J. and Jia, Z. (2024). Characterization of potential plastic-degradation enzymes from marine bacteria. *ACS omega*, 9(29):32185–32192.

Joshi, P., Banerjee, S., Hu, X., Khade, P. M., and Friedberg, I. (2023). Gothresher: a program to remove annotation biases from protein function annotation datasets. *Bioinformatics*, 39(1):btad048.

Kapinusova, G., Lopez Marin, M. A., and Uhlik, O. (2023). Reaching unreachables: Obstacles and successes of microbial cultivation and their reasons. *Frontiers in Microbiology*, 14:1089630.

Katz, L., Chen, Y. Y., Gonzalez, R., Peterson, T. C., Zhao, H., and Baltz, R. H. (2018). Synthetic biology advances and applications in the biotechnology industry: a perspective. *Journal of Industrial Microbiology and Biotechnology*, 45(7):449–461.

Khan, Z. U., Pi, D., Khan, I. A., Nawaz, A., Ahmad, J., and Hussain, M. (2021). pisaac: Extended notion of saac feature selection novel method for discrimination of enzymes model using different machine learning algorithm.

Khurana, S., Rawi, R., Kunji, K., Chuang, G.-Y., Bensmail, H., and Mall, R. (2018). Deepsol: a deep learning framework for sequence-based protein solubility prediction. *Bioinformatics*, 34(15):2605–2613.

Kim, G. B., Kim, J. Y., Lee, J. A., Norsigian, C. J., Palsson, B. O., and Lee, S. Y. (2023). Functional annotation of enzyme-encoding genes using deep learning with transformer layers. *Nature Communications*, 14(1):7370.

Kroll, A., Ranjan, S., Engqvist, M. K., and Lercher, M. J. (2023). A general model to predict small molecule substrates of enzymes based on machine and deep learning. *Nature communications*, 14(1):2787.

Kroll, A., Ranjan, S., and Lercher, M. J. (2024). A multimodal transformer network for protein-small molecule interactions enhances predictions of kinase inhibition and enzyme-substrate relationships. *PLOS Computational Biology*, 20(5):e1012100.

Kulmanov, M., Guzmán-Vega, F. J., Duek Roggli, P., Lane, L., Arold, S. T., and Hoehndorf, R. (2024). Protein function prediction as approximate semantic entailment. *Nature Machine Intelligence*, 6(2):220–228.

Kumar, S., Guruparan, D., Aaron, P., Telajan, P., Mahadevan, K., Davagandhi, D., and Yue, O. X. (2023). Deep learning in computational biology: Advancements, challenges, and future outlook. *arXiv preprint arXiv:2310.03086*.

Landeta, C., Medina-Ortiz, D., Escobar, N., Valdez, I., González-Troncoso, M. P., Álvares-Saravia, D., Aldridge, J., Gómez, C., and Lienqueo, M. E. (2024). Integrative workflows for the characterization of hydrophobin and cerato-platanin in the marine fungus paradendryphiella salina. *Archives of Microbiology*, 206(9):385.

Landwehr, G. M., Bogart, J. W., Magalhaes, C., Hammarlund, E. G., Karim, A. S., and Jewett, M. C. (2025). Accelerated enzyme engineering by machine-learning guided cell-free expression. *Nature Communications*, 16(1):865.

Li, F., Yuan, L., Lu, H., Li, G., Chen, Y., Engqvist, M. K. M., Kerkhoven, E. J., and Nielsen, J. (2022a). Deep learning-based kcat prediction enables improved enzyme-constrained model reconstruction. *Nature Catalysis*, 5(8):662–672.

Li, G., Rabe, K. S., Nielsen, J., and Engqvist, M. K. M. (2019). Machine learning applied to predicting microorganism growth temperatures and enzyme catalytic optima. *ACS Synthetic Biology*, 8(6):1411–1420.

Li, J. and Jiang, X. (2021). Mol-bert: An effective molecular representation with bert for molecular property prediction. *Wireless Communications and Mobile Computing*, 2021(1).

Li, X., Xiong, H., Li, X., Wu, X., Zhang, X., Liu, J., Bian, J., and Dou, D. (2022b). Interpretable deep learning: Interpretation, interpretability, trustworthiness, and beyond. *Knowledge and Information Systems*, 64(12):3197–3234.

Liaqat, S., Dashtipour, K., Arshad, K., Assaleh, K., and Ramzan, N. (2021). A hybrid posture detection framework: Integrating machine learning and deep neural networks. *IEEE Sensors Journal*, 21(7):9515–9522.

Libbrecht, M. W. and Noble, W. S. (2015). Machine learning applications in genetics and genomics. *Nature Reviews Genetics*, 16(6):321–332.

Lin, H., Chen, W., and Ding, H. (2013). Acalpred: A sequence-based tool for discriminating between acidic and alkaline enzymes. *PLoS ONE*, 8(10):e75726.

Longwell, C. K., Labanieh, L., and Cochran, J. R. (2017). High-throughput screening technologies for enzyme engineering. *Current Opinion in Biotechnology*, 48:196–202. Chemical biotechnology • Pharmaceutical biotechnology.

Madani, M., Lin, K., and Tarakanova, A. (2021). Dsressol: A sequence-based solubility predictor created with dilated squeeze excitation residual networks. *International Journal of Molecular Sciences*, 22(24):13555.

Mao, Y., Xu, W., Shun, Y., Chai, L., Xue, L., Yang, Y., and Li, M. (2025). A multimodal model for protein function prediction. *Scientific Reports*, 15(1):10465.

Markel, U., Essani, K. D., Besirlioglu, V., Schiffels, J., Streit, W. R., and Schwaneberg, U. (2020). Advances in ultrahigh-throughput screening for directed enzyme evolution. *Chemical Society Reviews*, 49(1):233–262.

Markus, B., Andreas, K., Arkadij, K., Stefan, L., Gustav, O., Elina, S., and Radka, S. (2023). Accelerating biocatalysis discovery with machine learning: a paradigm shift in enzyme engineering, discovery, and design. *ACS catalysis*, 13(21):14454–14469.

Martiny, H.-M., Armenteros, J. J. A., Johansen, A. R., Salomon, J., and Nielsen, H. (2021). Deep protein representations enable recombinant protein expression prediction. *Computational Biology and Chemistry*, 95:107596.

Medina-Ortiz, D., Alvarez-Saravia, D., Soto-García, N., Sandoval-Vargas, D., Aldridge, J., Rodríguez, S., Andrews, B., Asenjo, J. A., and Daza, A. (2025). Protein language models accelerate the discovery of plastic-degrading enzymes. *bioRxiv*.

Medina-Ortiz, D., Cabas-Mora, G., Moya-Barría, I., Soto-Garcia, N., and Uribe-Paredes, R. (2024a). Rudeus, a machine learning classification system to study dna-binding proteins. *bioRxiv*, pages 2024–02.

Medina-Ortiz, D., Contreras, S., Amado-Hinojosa, J., Torres-Almonacid, J., Asenjo, J. A., Navarrete, M., and Olivera-Nappa, Á. (2022). Generalized property-based encoders and digital signal processing facilitate predictive tasks in protein engineering. *Frontiers in Molecular Biosciences*, 9:898627.

Medina-Ortiz, D., Contreras, S., Fernández, D., Soto-García, N., Moya, I., Cabas-Mora, G., and Olivera-Nappa, Á. (2024b). Protein language models and machine learning facilitate the identification of antimicrobial peptides. *International Journal of Molecular Sciences*, 25(16):8851.

Medina-Ortiz, D., Khalifeh, A., Anvari-Kazemabad, H., and Davari, M. D. (2024c). Interpretable and explainable predictive machine learning models for data-driven protein engineering. *Biotechnology Advances*, page 108495.

Mellor, J., Grigoras, I., Carbonell, P., and Faulon, J.-L. (2016). Semisupervised gaussian process for automated enzyme search. *ACS Synthetic Biology*, 5(6):518–528.

Merino, G. A., Saidi, R., Milone, D. H., Stegmayer, G., and Martin, M. J. (2022). Hierarchical deep learning for predicting go annotations by integrating protein knowledge. *Bioinformatics*, 38(19):4488–4496.

Minton, A. P. (2001). The influence of macromolecular crowding and macromolecular confinement on biochemical reactions in physiological media *. *Journal of Biological Chemistry*, 276(14):10577–10580.

Mou, Z., Eakes, J., Cooper, C. J., Foster, C. M., Standaert, R. F., Podar, M., Doktycz, M. J., and Parks, J. M. (2020). Machine learning-based prediction of enzyme substrate scope: Application to bacterial nitrilases. *Proteins: Structure, Function, and Bioinformatics*, 89(3):336–347.

Nam, H., Lewis, N. E., Lerman, J. A., Lee, D.-H., Chang, R. L., Kim, D., and Palsson, B. O. (2012). Network context and selection in the evolution to enzyme specificity. *Science*, 337(6098):1101–1104.

Nikolados, E.-M., Wongprommoon, A., Aodha, O. M., Cambray, G., and Oyarzún, D. A. (2022). Accuracy and data efficiency in deep learning models of protein expression. *Nature Communications*, 13(1):7755.

Norton-Baker, B., Denton, M. C., Murphy, N. P., Fram, B., Lim, S., Erickson, E., Gauthier, N. P., and Beckham, G. T. (2024). Enabling high-throughput enzyme discovery and engineering with a low-cost, robot-assisted pipeline. *Scientific Reports*, 14(1):14449.

Notin, P., Kollasch, A., Ritter, D., Van Niekerk, L., Paul, S., Spinner, H., Rollins, N., Shaw, A., Orenbuch, R., Weitzman, R., et al. (2023). Proteingym: Large-scale benchmarks for protein fitness prediction and design. *Advances in Neural Information Processing Systems*, 36:64331–64379.

Oberg, N., Zallot, R., and Gerlt, J. A. (2023). Efi-est, efi-gnt, and efi-cgfp: enzyme function initiative (efi) web resource for genomic enzymology tools. *Journal of molecular biology*, 435(14):168018.

Pan, T., Li, C., Bi, Y., Wang, Z., Gasser, R. B., Purcell, A. W., Akutsu, T., Webb, G. I., Imoto, S., and Song, J. (2023). Pfresgo: an attention mechanism-based deep-learning approach for protein annotation by integrating gene ontology inter-relationships. *Bioinformatics*, 39(3).

Paysan-Lafosse, T., Blum, M., Chuguransky, S., Grego, T., Pinto, B. L., Salazar, G. A., Bileschi, M. L., Bork, P., Bridge, A., Colwell, L., et al. (2023). Interpro in 2022. *Nucleic acids research*, 51(D1):D418–D427.

Pei, H., Li, J., Ma, S., Jiang, J., Li, M., Zou, Q., and Lv, Z. (2023). Identification of thermophilic proteins based on sequence-based bidirectional representations from transformer-embedding features. *Applied Sciences*, 13(5):2858.

Probst, D., Manica, M., Nana Teukam, Y. G., Castrogiovanni, A., Paratore, F., and Laino, T. (2022). Biocatalysed synthesis planning using data-driven learning. *Nature Communications*, 13(1).

Qiu, S., Hu, B., Zhao, J., Xu, W., and Yang, A. (2025). Seq2topt: a sequence-based deep learning predictor of enzyme optimal temperature. *Briefings in Bioinformatics*, 26(2).

Radivojac, P., Clark, W. T., Oron, T. R., Schnoes, A. M., Wittkop, T., Sokolov, A., Graim, K., Funk, C., Verspoor, K., Ben-Hur, A., et al. (2013). A large-scale evaluation of computational protein function prediction. *Nature methods*, 10(3):221–227.

Richardson, L., Allen, B., Baldi, G., Beracochea, M., Bileschi, M. L., Burdett, T., Burgin, J., Caballero-Pérez, J., Cochrane, G., Colwell, L. J., et al. (2023). Mgnify: the microbiome sequence data analysis resource in 2023. *Nucleic acids research*, 51(D1):D753–D759.

Rives, A., Meier, J., Sercu, T., Goyal, S., Lin, Z., Liu, J., Guo, D., Ott, M., Zitnick, C. L., Ma, J., and Fergus, R. (2021). Biological structure and function emerge from scaling unsupervised learning to 250 million protein sequences. *Proceedings of the National Academy of Sciences*, 118(15).

Robinson, P. K. (2015). Enzymes: principles and biotechnological applications. *Essays in biochemistry*, 59:1.

Robinson, S. L., Piel, J., and Sunagawa, S. (2021). A roadmap for metagenomic enzyme discovery. *Natural Product Reports*, 38(11):1994–2023.

Robinson, S. L., Terlouw, B. R., Smith, M. D., Pidot, S. J., Stinear, T. P., Medema, M. H., and Wackett, L. P. (2020). Global analysis of adenylate-forming enzymes reveals β-lactone biosynthesis pathway in pathogenic nocardia. *Journal of Biological Chemistry*, 295(44):14826–14839.

Samusevich, R., Hebra, T., Bushuiev, R., Engst, M., Kulhánek, J., Bushuiev, A., Smith, J. D., Čalounová, T., Smrčková, H., Molineris, M., et al. (2024). Structure-enabled enzyme function prediction unveils elusive terpenoid biosynthesis in archaea. *BioRxiv*, pages 2024–01.

Sanderson, T., Bileschi, M. L., Belanger, D., and Colwell, L. J. (2023). ProteInfer, deep neural networks for protein functional inference. *eLife*, 12.

Sapoval, N., Aghazadeh, A., Nute, M. G., Antunes, D. A., Balaji, A., Baraniuk, R., Barberan, C., Dannenfelser, R., Dun, C., Edrisi, M., et al. (2022). Current progress and open challenges for applying deep learning across the biosciences. *Nature Communications*, 13(1):1728.

Scherer, M., Fleishman, S. J., Jones, P. R., Dandekar, T., and Bencurova, E. (2021). Computational enzyme engineering pipelines for optimized production of renewable chemicals. *Frontiers in bioengineering and biotechnology*, 9:673005.

Scherlach, K. and Hertweck, C. (2021). Mining and unearthing hidden biosynthetic potential. *Nature Communications*, 12(1):3864.

Sharma, A., Bagler, G., and Bera, D. (2019). Supervised learning of protein thermal stability using sequence mining and distribution statistics of network centrality. *BioRxiv*, page 777177.

Shi, Z., Yuan, Q., Wang, R., Li, H., Liao, X., and Ma, H. (2022). Ecrecer: Enzyme commission number recommendation and benchmarking based on multiagent dual-core learning. *arXiv preprint arXiv:2202.03632*.

Shinde, R. M., Gupta, S., Kuma, D., and Ansar, M. J. (2024). Enzyme annotation and pathway reconstruction with machine learning. In *Genomic Intelligence*, pages 119–136. CRC Press.

Siedhoff, N. E., Schwaneberg, U., and Davari, M. D. (2020). Machine learning-assisted enzyme engineering. *Methods in enzymology*, 643:281–315.

Silva, D., Rodrigues, C. F., Lorena, C., Borges, P. T., and Martins, L. O. (2023). Biocatalysis for biorefineries: The case of dye-decolorizing peroxidases. *Biotechnology Advances*, 65:108153.

Steinegger, M. and Söding, J. (2017). Mmseqs2 enables sensitive protein sequence searching for the analysis of massive data sets. *Nature biotechnology*, 35(11):1026–1028.

Su, J., Zhang, F., Sun, W., Karuppiah, V., Zhang, G., Li, Z., and Jiang, Q. (2015). A new alkaline lipase obtained from the metagenome of marine sponge ircinia sp. *World Journal of Microbiology and Biotechnology*, 31(7):1093–1102.

Teufel, J., Leinweber, A., and Friederich, P. (2025). Improving counterfactual truthfulness for molecular property prediction through uncertainty quantification. *arXiv preprint arXiv:2504.02606*.

Thumuluri, V., Martiny, H.-M., Almagro Armenteros, J. J., Salomon, J., Nielsen, H., and Johansen, A. R. (2021). Netsolp: predicting protein solubility in escherichia coli using language models. *Bioinformatics*, 38(4):941–946.

Van Kempen, M., Kim, S. S., Tumescheit, C., Mirdita, M., Lee, J., Gilchrist, C. L., Söding, J., and Steinegger, M. (2024). Fast and accurate protein structure search with foldseek. *Nature biotechnology*, 42(2):243–246.

Vasina, M., Vanacek, P., Damborsky, J., and Prokop, Z. (2020). Exploration of enzyme diversity: High-throughput techniques for protein production and microscale biochemical characterization. In *Methods in Enzymology*, volume 643, pages 51–85. Elsevier.

Wang, H., Ren, Z., Sun, J., Chen, Y., Bo, X., Xue, J., Gao, J., and Ni, M. (2025a). Deeppfp: a multi-task-aware architecture for protein function prediction. *Briefings in Bioinformatics*, 26(1):bbae579.

Wang, J., Yang, Z., Chen, C., Yao, G., Wan, X., Bao, S., Ding, J., Wang, L., and Jiang, H. (2024). Mpek: a multitask deep learning framework based on pretrained language models for enzymatic reaction kinetic parameters prediction. *Briefings in Bioinformatics*, 25(5):bbae387.

Wang, X., Liu, Y., Du, Z., Zhu, M., Kaushik, A. C., Jiang, X., and Wei, D. (2021). Prediction of protein solubility based on sequence feature fusion and ddccnn. *Interdisciplinary Sciences: Computational Life Sciences*, 13(4):703–716.

Wang, Y., Cheng, L., Zhang, Y., Cao, Y., and Alghazzawi, D. (2025b). Dekp: a deep learning model for enzyme kinetic parameter prediction based on pretrained models and graph neural networks. *Briefings in Bioinformatics*, 26(2):bbaf187.

Wang, Z., Fan, J., Guo, R., Nguyen, T., Ji, H., and Liu, G. (2025c). Proteinzero: Self-improving protein generation via online reinforcement learning. *arXiv preprint arXiv:2506.07459*.

Wang, Z., Xie, D., Wu, D., Luo, X., Wang, S., Li, Y., Yang, Y., Li, W., and Zheng, L. (2025d). Robust enzyme discovery and engineering with deep learning using catapro. *Nature communications*, 16(1):2736.

Watanabe, N., Yamamoto, M., Murata, M., Vavricka, C. J., Ogino, C., Kondo, A., and Araki, M. (2022). Comprehensive machine learning prediction of extensive enzymatic reactions. *The Journal of Physical Chemistry B*, 126(36):6762–6770.

Xing, H., Cai, P., Liu, D., Han, M., Liu, J., Le, Y., Zhang, D., and Hu, Q.-N. (2024). High-throughput prediction of enzyme promiscuity based on substrate–product pairs. *Briefings in Bioinformatics*, 25(2).

Xu, M., Zhang, Z., Lu, J., Zhu, Z., Zhang, Y., Ma, C., Liu, R., and Tang, J. (2022). Peer: A comprehensive and multi-task benchmark for protein sequence understanding.

Yan, S. and Wu, G. (2019). Predictors for predicting temperature optimum in beta-glucosidases. *Journal of Biomedical Science and Engineering*, 12(08):414–426.

Yang, J., Li, F.-Z., and Arnold, F. H. (2024a). Opportunities and challenges for machine learning-assisted enzyme engineering. *ACS Central Science*, 10(2):226–241.

Yang, J., Mora, A., Liu, S., Wittmann, B. J., Anandkumar, A., Arnold, F. H., and Yue, Y. (2024b). Care: a benchmark suite for the classification and retrieval of enzymes.

Yang, M., Fehl, C., Lees, K. V., Lim, E.-K., Offen, W. A., Davies, G. J., Bowles, D. J., Davidson, M. G., Roberts, S. J., and Davis, B. G. (2018). Functional and informatics analysis enables glycosyltransferase activity prediction. *Nature Chemical Biology*, 14(12):1109–1117.

Yeo, J., Han, Y., Bordin, N., Lau, A. M., Kandathil, S. M., Kim, H., Levy Karin, E., Mirdita, M., Jones, D. T., Orengo, C., et al. (2025). Metagenomic-scale analysis of the predicted protein structure universe. *bioRxiv*, pages 2025–04.

Yu, H., Deng, H., He, J., Keasling, J. D., and Luo, X. (2023a). Unikp: a unified framework for the prediction of enzyme kinetic parameters. *Nature Communications*, 14(1).

Yu, H. and Luo, X. (2024). Thermofinder: A sequence-based thermophilic proteins prediction framework. *International Journal of Biological Macromolecules*, 270:132469.

Yu, T., Cui, H., Li, J. C., Luo, Y., Jiang, G., and Zhao, H. (2023b). Enzyme function prediction using contrastive learning. *Science*, 379(6639):1358–1363.

Yuan, Q., Xie, J., Xie, J., Zhao, H., and Yang, Y. (2023). Fast and accurate protein function prediction from sequence through pretrained language model and homology-based label diffusion. *Briefings in Bioinformatics*, 24(3).

Zhang, D., Xing, H., Liu, D., Han, M., Cai, P., Lin, H., Tian, Y., Guo, Y., Sun, B., Tian, Y., Aibo, W., and Hu, Q.-N. (2023). Deep learning enables rapid identification of mycotoxin-degrading enzymes. *ChemRxiv*.

Zhang, G., Li, H., and Fang, B. (2009). Discriminating acidic and alkaline enzymes using a random forest model with secondary structure amino acid composition. *Process Biochemistry*, 44(6):654–660.

Zhang, H., Perez-Garcia, P., Dierkes, R. F., Applegate, V., Schumacher, J., Chibani, C. M., Sternagel, S., Preuss, L., Weigert, S., Schmeisser, C., Danso, D., Pleiss, J., Almeida, A., Höcker, B., Hallam, S. J., Schmitz, R. A., Smits, S. H. J., Chow, J., and Streit, W. R. (2022a). The bacteroidetes aequorivita sp. and kaistella jeonii produce promiscuous esterases with pet-hydrolyzing activity. *Frontiers in Microbiology*, 12.

Zhang, Y., Bian, B., and Okumura, M. (2025). Hyena architecture enables fast and efficient protein language modeling. *IMetaOmics*, 2(1):e45.

Zhang, Y., Guan, F., Xu, G., Liu, X., Zhang, Y., Sun, J., Yao, B., Huang, H., Wu, N., and Tian, J. (2022b). A novel thermophilic chitinase directly mined from the marine metagenome using the deep learning tool preoptem. *Bioresources and Bioprocessing*, 9(1).

Zhao, J., Yan, W., and Yang, Y. (2023). Deeptp: A deep learning model for thermophilic protein prediction. *International Journal of Molecular Sciences*, 24(3):2217.

Zhapa-Camacho, F., Tang, Z., Kulmanov, M., and Hoehndorf, R. (2024). Predicting protein functions using positive-unlabeled ranking with ontology-based priors. *Bioinformatics*, 40(Supplement_1):i401–i409.

Zhou, J. and Huang, M. (2024). Navigating the landscape of enzyme design: from molecular simulations to machine learning. *Chemical Society Reviews*, 53(16):8202–8239.

Zhou, L., Tao, C., Shen, X., Sun, X., Wang, J., and Yuan, Q. (2024). Unlocking the potential of enzyme engineering via rational computational design strategies. *Biotechnology Advances*, page 108376.

# Supplementary Material

# Machine Learning-Driven Enzyme Mining: Opportunities, Challenges, and Future Perspectives

Yanzi Zhang[1§], Felix Moorhoff[2§], Sizhe Qiu[3], Wenjuan Dong[1], David Medina-Ortiz[2]

Jing Zhao[1,*] and Mehdi D. Davari[2,*]

[1] State Key Laboratory of Biocatalysis and Enzyme Engineering, Hubei Key Laboratory of Industrial Biotechnology, School of Life Sciences, Hubei University, Wuhan, 430062, China

[2] Department of Bioorganic Chemistry, Leibniz Institute of Plant Biochemistry, Weinberg 3, 06120, Halle, Germany

[3] Department of Engineering Science, University of Oxford, OX1 3PJ, United Kingdom

[§]Y. Z. and F. M. contributed equally to this work.

*Corresponding authors:

Dr. Jing Zhao, E-mail: zhaojing@hubu.edu.cn; zhaojing802@hotmail.com

Dr. Mehdi D. Davari, E-mail: mehdi.davari@ipb-halle.de

# Table of Contents

# List of Abbreviations

| Full name | Abbreviations | Full name | Abbreviations |
|---|---|---|---|
| Artificial Neural Network | ANN | Language Models | LMs |
| Bidirectional Encoder Representations from Transformers | BERT | Linear Regression | LR |
| Bidirectional Long Short-Term Memory Network | BiLSTM | Long Short-Term Memory | LSTM |
| Customized Gate Control | CGC | Machine Learning | ML |
| Convolutional Neural Network | CNN | Melting Temperature | $T_m$ |
| Deep Convolutional Neural Networks | D-CNNs | Multi-layer Perceptron | MLP |
| Deep Neural Network | DNN | Mixed Logistic Regression | MLR |
| Decision Tree | DT | Not Applicable or Not Available | N/A |
| Enzyme Commission | EC | Natural Language Processing | NLP |
| eXtreme Gradient Boosting | XGBoost | Neural Network | NN |
| Extreme Learning Machine | ELM | Optimal pH | $pH_{opt}$ |
| Evolutionary Scale Modeling | ESM | Optimal Temperature | $T_{opt}$ |
| Fully Connected Neural Network | FCNN | Protein Family Database | Pfam |
| Feedforward Neural Network | FNN | Probabilistic Neural Network | PNN |
| Genetic Algorithm | GA | Random Forest | RF |
| Graph Attention Networks | GAT | Recurrent Neural Network | RNN |
| Gradient Boosting Decision Tree | GBDT | Scoring Card Method | SCM |
| Gradient Boosting Machine | GBM | Sequence | Seq |
| Graph Convolutional Network | GCN | Support Vector Classification | SVC |
| Gene Ontology | GO | Support Vector Machine | SVM |
| Gaussian Process | GP | Support Vector Regression | SVR |
| Gated Recurrent Unit | GRU | Temperature | Temp |
| Hidden Markov Models | HMMs | | |
| k-nearest neighbor classification | kNN | | |

**Table S1. Selected ML tools for predicting enzyme EC numbers and GO terms (2014-Present).**

| Tool | ML method | Input | Output | Availability | Year | Ref |
|------|-----------|-------|--------|--------------|------|-----|
| ECPred | Ensemble (kNN, SVM) | Seq | EC number | http://cansyl.metu.edu.tr/ECPred.html | 2018 | 1 |
| goPredSim | LMs | Seq | GO terms | https://embed.protein.properties/ | 2021 | 2 |
| EPP-HMCNF | Hierarchical multi-label neural network | Enzyme-substrate interactions | EC numbers | Downloadable | 2021 | 3 |
| DeepFRI | GCN, LSTM | Seq | EC numbers, GO terms | https://beta.deepfri.flatironinstitute.org/ | 2021 | 4 |
| BENZ WS | HMMs, Pfam mode | Seq | EC numbers | https://benzdb.biocomp.unibo.it/ | 2021 | 5 |
| mlHECNet | CNN, RNN | Seq | EC numbers | http://hecnet.cbrlab.org/ | 2021 | 6 |
| ABLE | Attention Mechanism, LSTM | Seq | EC numbers | Downloadable | 2021 | 7 |
| ProtCNN | CNN | Seq | Enzyme function prediction | N/A | 2022 | 8 |
| SEP-DNN model | DNN | Seq | Enzyme reaction predictions | N/A | 2022 | 9 |
| Pandya et al. | Ensemble (SVM, RF etc.) | Seq | Enzyme family classification | N/A | 2022 | 10 |
| ProPythia | SVM, DNNs etc. | Seq | EC numbers | Downloadable | 2022 | 11 |
| DeeProtGO | FNN | Seq | GO terms | Downloadable | 2022 | 12 |
| PFresGO | Attention mechanism | Seq | GO terms | Downloadable | 2023 | 13 |
| ProteInfer | CNN | Seq | EC numbers, GO terms | Downloadable | 2023 | 14 |
| CLEAN | Contrastive learning | Seq | EC numbers | N/A | 2023 | 15 |
| EnzymeNet | ResNets | Seq | EC numbers | Downloadable | 2023 | 16 |
| ProtEC | Transformer | Seq | EC numbers | N/A | 2023 | 17 |
| SPROF-GO | Ensemble (ProtTrans, Self-Attention, MLP etc.) | Seq | GO terms | http://bio-web1.nscc-gz.cn/app/sprof-go | 2023 | 18 |
| EZYDeep | CNN | Seq | EC numbers | N/A | 2023 | 19 |
| Fernández et al. | CNN | Seq | EC numbers | Downloadable | 2023 | 20 |

| EnzBert | Transformer | Seq | EC numbers | Downloadable | 2023 | 21 |
| PARSE | COLLAPSE | Structure | Enzyme function prediction | Downloadable | 2023 | 22 |
| DeepECtransformer | BERT, self-attention | Seq | EC numbers | N/A | 2023 | 23 |
| ECRECer | GRU, attention mechanism | Seq | EC numbers | http://ecrecer.biodesign.ac.cn | 2023 | 24 |
| HiFi-NN | Contrastive learning | Seq | EC numbers | N/A | 2023 | 25 |
| ECPICK | CNN, hierarchical layers | Seq | EC numbers | http://ecpick.dataxlab.org | 2023 | 26 |
| MVDINET | Ensemble (CNN, MLP) | Seq | EC numbers | Downloadable | 2024 | 27 |

**Table S2. Selected ML tools for predicting enzyme substrate specificity (2014-Present).**

| Tool | ML method | Input | Output | Scope[1] | Availability | Year | Ref |
|---|---|---|---|---|---|---|---|
| Pertusi et al. | SVM | Seq, Substrate | Score | MenD & Car & AAEH & HAPMO | N/A | 2017 | 28 |
| GT-predict | DT | Seq | Transformations, functional annotation | Glycosyltransferase | N/A | 2018 | 29 |
| Mou et al. | RF | Seq, Substrate | Active/ inactive | Bacterial Nitrilase | Downloadable | 2020 | 30 |
| pNPred | RF | Seq | Active/ inactive | Thiolases | z.umn.edu/thiolases | 2020 | 31 |
| Adenylpred | RF | Seq | substrate specificity | ANL enzyme | https://srobinson.shinyapps.io/ AdenylPred/ | 2020 | 32 |
| EnZymClass | SVM | Seq | Categories | Plant acyl-ACP thioesterase | Downloadable | 2022 | 33 |
| ProSmith | Transformer, GBM | Seq, Substrate | Score | General | Downloadable | 2023 | 34 |
| ESP | GBDT | Seq, Substrate | Score | General | https://esp.cs.hhu.de | 2023 | 35 |
| PGCN | GCN | Seq, Substrate | Score | Protease | N/A | 2023 | 36 |
| Rappoport et al. | XGBoost | Structure, Substrate | Cofactor & substrate structural class preference | SDRs & SAM-MTase | N/A | 2023 | 37 |
| GASP | RF | Seq, Substrate | Score | Family 1 Glycosyltransferase | Downloadable | 2023 | 38 |

[1] MenD & Car & AAEH & HAPMO: 2-succinyl-5-enolpyruvyl-6-hydroxy-3-cyclohexene-1-carboxylic acid synthase & carboxylic acid reductase & amino acid ester hydrolase & 4-hydroxyacetophenone monooxygenase; SDRs & SAM-Mtase: short-chain dehydrogenase/reductases & S-adenosylmethionine-dependent methyltransferases.

**Table S3. Selected ML tools for predicting enzyme kinetic parameters (2014-Present).**

| Tool | ML method | Input | Output | Availability | Year | Ref |
|------|-----------|-------|--------|--------------|------|-----|
| Mellor et al. | GP model | Reaction signatures, Seq | $K_m$ | N/A | 2016 | 39 |
| Heckmann et al. | Ensemble learning of ElasticNet, RF, ANN | Seq, Structure, pH, Temp etc. | $k_{cat}$ | N/A | 2018 | 40 |
| Kroll et al. | Deep representation, GNN, XGBoost | Seq, Substrate | $K_m$ | Downloadable | 2021 | 41 |
| DLKcat | GNN, attention CNN | Seq, Substrate | $k_{cat}$ | Downloadable | 2022 | 42 |
| TurNuP | Deep representation and XGBoost | Seq, Substrate | $k_{cat}$ | https://turnup.cs.hhu.de | 2023 | 43 |
| UniKP | Deep representation, Extra Trees model | Seq, Substrate | $K_m$, $k_{cat}$, $K_m/k_{cat}$ | Downloadable | 2023 | 44 |
| EF-UniKP | Deep representation, Extra Trees model | Seq, Substrate, Temp, pH | $k_{cat}$ | Downloadable | 2023 | 44 |
| GELKcat | Attention CNN and GAT | Seq, Substrate | $k_{cat}$ | N/A | 2023 | 45 |
| ProSmith | Deep representation, BERT, XGBoost | Seq, Substrate | $K_m$ | Downloadable | 2023 | 34 |
| DeepEnzyme | Attention GCN | Seq, Structure, Substrate | $k_{cat}$ | Downloadable | 2024 | 46 |
| DLTKcat | Attention GNN, CNN, bi-directional attention mechanism | Seq, Substrate, Temp | $k_{cat}$ | Downloadable | 2024 | 47 |
| CatPred | Deep representation, self-attention and GNN | Seq, Structure, Substrate | $K_m$, $k_{cat}$ | https://tiny.cc/catpred | 2024 | 48 |
| MPEK | Deep representation (ProtT5+Mole-BERT) and CGC framework | Seq, Substrate, Temp, pH, organismal information | $K_m$, $k_{cat}$ | http://mathtc.nscc-tj.cn/mpek | 2024 | 49 |

**Table S4. Selected ML tools for predicting enzyme thermostability (2014-Present).**

| Tool | ML method | Input | Output | Availability | Year | Ref |
|------|-----------|-------|--------|--------------|------|-----|
| Chakravorty et al. | SVM | Seq | Thermal stable or not | N/A | 2016 | 50 |
| Rezaeenour et al. | ELM | Seq | Thermal stable or not | N/A | 2016 | 51 |
| DeepET | CNN | Seq | Tm | Downloadable | 2022 | 52 |
| ColGen | BiLSTM+FCNN | Seq | Tm | N/A | 2022 | 53 |
| ProTstab2 | GBM | Seq | Tm | http://structure.bmc.lu.se/ProTstab2/ | 2022 | 54 |
| DeepSTABp | Transformer | Seq | Tm | https://csb-deepstabp.bio.rptu.de | 2023 | 55 |
| DeepTM | GCN | Seq | Tm | http://deeptm.top/index.html | 2023 | 56 |

**Table S5. Selected ML tools for predicting enzyme thermophilicity (2014-Present).** Input types of these tools are sequences.

| Tool | ML method | Output | Availability | Year | Ref |
|------|-----------|--------|--------------|------|-----|
| Wang and Li | GA and MLR | Thermophilic or not | N/A | 2014 | 57 |
| Fan et al. | SVM | Thermophilic or mesophilic | N/A | 2016 | 58 |
| Tang et al. | Fragment based classification | Thermophilic or not | N/A | 2017 | 59 |
| Protein_Classification _BIO699 | RF | Thermophilic or mesophilic | Downloadable | 2019 | 60 |
| Li et al. | Voting algorithm | Thermophilic or not | N/A | 2019 | 61 |
| Feng et al. | SVM | Thermophilic or not | N/A | 2020 | 62 |
| SCMTPP | SCM | Thermophilic or not | http://pmlabstack.pythonanywhere.com/SCMTPP | 2021 | 63 |
| TAXyl | RF | non-thermophilic, thermophilic, or hyper-thermophilic | arimees.com | 2021 | 64 |
| TMPpred | SVM | Thermophilic or not | http://112.124.26.17:7000/TMPpred/index.html | 2022 | 65 |
| iThermo | MLP | Thermophilic or not | Downloadable | 2022 | 66 |
| SAPPHIRE | Stacking ensemble learning | Thermophilic or not | Downloadable | 2022 | 67 |
| ThermoProt | SVM | PM, MT, TH, MTH | Downloadable | 2022 | 68 |
| BertThermo | Deep representation and LR | Thermophilic or not | Downloadable | 2023 | 69 |
| ProLaTherm | Deep representation and DNN | Thermophilic or not | Downloadable | 2023 | 70 |
| DeepTP | CNN, BiLSTM and self-attention mechanism | Thermophilic or mesophilic | http://www.YangLab-MI.org.cn/DeepTP | 2023 | 71 |
| Wan et al. | Decision tree | Thermophilic or not | N/A | 2023 | 72 |
| DeepPPThermo | DNN, deep representation, BiLSTM and attention mechanism | Thermophilic or mesophilic | Downloadable | 2024 | 73 |
| ThermoFinder | Deep representation and ensemble learning | Thermophilic or not | Downloadable | 2024 | 74 |

**Table S6. Selected ML tools for predicting enzyme acidophily and alkaliphily (2009-Present).**

| Tool | ML method | Input | Output | Availability | Year | Ref |
|------|-----------|-------|--------|--------------|------|-----|
| RF model | RF | Seq | Acidic or alkaline | N/A | 2009 | 75 |
| SVM model | SVM | Seq | Acidic or alkaline | http://wlxy.imu.edu.cn/college/biostation/fuwu/enzymes/index.asp | 2013 | 76 |
| AcalPred | SVM | Seq | Acidic or alkaline | http://lin.uestc.edu.cn/server/AcalPred | 2013 | 77 |
| PNN model | PNN | Seq | Acidic or alkaline | N/A | 2014 | 78 |
| SVM model2 | SVM | Seq | Acidic or alkaline | N/A | 2019 | 79 |
| piSAAC | DNN | Seq | Acidic or alkaline | N/A | 2020 | 80 |

**Table S7. Selected ML tools for predicting enzyme optimal temperature and pH (2014-Present).**

| Tool | ML method | Scope | Output | Availability | Year | Ref |
|------|-----------|-------|--------|--------------|------|-----|
| Chu et al. | SVR | Only for $\beta$-agarases | $T_{opt}$ | N/A | 2016 | 81 |
| Yan and Wu | ANN | Only for $\beta$-Glucosidases | $T_{opt}$, $pH_{opt}$ | N/A | 2019 | 82, 83 |
| TOME | SVR | General | $T_{opt}$ | Downloadable | 2019 | 84 |
| TOMER | Ensemble averaging | General | $T_{opt}$ | Downloadable | 2020 | 85 |
| MeTarEnz | SVR and RF | Only for lipases | $T_{opt}$, $pH_{opt}$ | Downloadable | 2022 | 86 |
| Preoptem | One-hot encoding and CNN | General | $T_{opt}$ | N/A | 2022 | 87 |
| EpHod | Deep representation, lightweight attention and CNN | General | $pH_{opt}$ | Downloadable | 2023 | 88 |
| Seq2Topt | ESM-2 and multi-head attention | General | $T_{opt}$ | Downloadable | 2024 | 89 |
| OphPred | ESM-2 and XGBoost | General | $pH_{opt}$ | Downloadable | 2024 | 90 |

**Table S8. Selected ML tools for predicting enzyme solubility (2014-Present).**

| Tool | ML method | Input | Output | Availability | Year | Ref |
|------|-----------|-------|--------|--------------|------|-----|
| ccSol omics | SVM | Seq | Solubility score, Soluble/insoluble | http://s.tartaglialab.com/page/ccsol_group | 2014 | 91 |
| Protein-Sol | LR | Seq | Scaled solubility value | http://protein-sol.manchester.ac.uk | 2017 | 92 |
| DeepSol | CNN | Seq | Solubility score | Downloadable | 2018 | 93 |
| PaRSnIP | GBM | Seq | Solubility score | Downloadable | 2018 | 94 |
| SolXplain | GBM | Seq | Solubility score | N/A | 2019 | 95 |
| ProGAN | DNN | Seq | Solubility score | Downloadable | 2019 | 96 |
| Han et al. | SVM | Seq | soluble or insoluble | Downloadable | 2019 | 97 |
| SoDoPE | LR | Seq | Solubility score, Flexibility | https://tisigner.com/sodope | 2020 | 98 |
| SKADE | NN | Seq | Solubility score | Downloadable | 2020 | 99 |
| SOLart | RF | Structure | Scaled solubility value, soluble/insoluble | http://babylone.ulb.ac.be/SOLART/ | 2020 | 100 |
| UniRep-RF | RNN | Seq | Solubility score | Downloadable | 2021 | 101 |
| EPSOL | MLP | Seq | Solubility score | Downloadable | 2021 | 102 |
| SoluProt | GBM | Seq | Solubility score | https://loschmidt.chemi.muni.cz/soluprot/ | 2021 | 103 |
| GraphSol | GCN | Seq | Solubility score | Downloadable | 2021 | 104 |
| DSResSol | D-CNNs | Seq | Solubility score, Expression score | https://tgs.uconn.edu/dsres_sol | 2021 | 105 |
| NetSolP | Transformer | Seq | Solubility score | https://services.healthtech.dtu.dk/service.php?NetSolP | 2022 | 106 |
| HybridGCN | GCN | Seq | Solubility score | Downloadable | 2023 | 107 |

| PPSol | GCN | Seq | Solubility score | Downloadable | 2023 | 108 |
| DeepSoluE | LSTM | Seq | Probability, soluble/insoluble | http://lab.malab.cn/~wangchao/softs/DeepSoluE/ | 2023 | 109 |
| DDcCNN | CNN | Seq | Solubility score | Downloadable | 2023 | 110 |
| RPPSP | RF | Seq | Solubility score | Downloadable | 2023 | 111 |

# REFERENCES

(1) Dalkiran, A.; Rifaioglu, A. S.; Martin, M. J.; Cetin-Atalay, R.; Atalay, V.; Doğan, T., ECPred: a tool for the prediction of the enzymatic functions of protein sequences based on the EC nomenclature. *BMC Bioinf.* **2018**, *19* (1), 334.

(2) Littmann, M.; Heinzinger, M.; Dallago, C.; Olenyi, T.; Rost, B., Embeddings from deep learning transfer GO annotations beyond homology. *Sci. Rep.* **2021**, *11* (1), 1160.

(3) Visani, G. M.; Hughes, M. C.; Hassoun, S., Enzyme Promiscuity Prediction Using Hierarchy-Informed Multi-Label Classification. *Bioinformatics* **2021**, *37* (14), 2017-2024.

(4) Gligorijević, V.; Renfrew, P. D.; Kosciolek, T.; Leman, J. K.; Berenberg, D.; Vatanen, T.; Chandler, C.; Taylor, B. C.; Fisk, I. M.; Vlamakis, H.; Xavier, R. J.; Knight, R.; Cho, K.; Bonneau, R., Structure-based protein function prediction using graph convolutional networks. *Nat. Commun.* **2021**, *12* (1), 3168.

(5) Baldazzi, D.; Savojardo, C.; Martelli, P. L.; Casadio, R., BENZ WS: the Bologna ENZyme Web Server for four-level EC number annotation. *Nucleic Acids Res.* **2021**, *49* (W1), W60-W66.

(6) Khan, K. A.; Memon, S. A.; Naveed, H., A hierarchical deep learning based approach for multi-functional enzyme classification. *Protein Sci.* **2021**, *30* (9), 1935-1945.

(7) Nallapareddy, M. V.; Dwivedula, R., ABLE: Attention based learning for enzyme classification. *Comput. Biol. Chem.* **2021**, *94*, 107558.

(8) Bileschi, M. L.; Belanger, D.; Bryant, D. H.; Sanderson, T.; Carter, B.; Sculley, D.; Bateman, A.; DePristo, M. A.; Colwell, L. J., Using deep learning to annotate the protein universe. *Nat. Biotechnol.* **2022**, *40* (6), 932-937.

(9) Watanabe, N.; Yamamoto, M.; Murata, M.; Vavricka, C. J.; Ogino, C.; Kondo, A.; Araki, M., Comprehensive Machine Learning Prediction of Extensive Enzymatic Reactions. *J. Phys. Chem. B* **2022**, *126* (36), 6762-6770.

(10) Pandya, D. D.; Jadeja, A.; Degadwala, S.; Vyas, D. Ensemble Learning based Enzyme Family Classification using n-gram Feature. In Proceedings of the 6th International Conference on Intelligent Computing and Control Systems (ICICCS); 25-27 May **2022**; pp 1386-1392.

(11) Sequeira, A. M.; Lousa, D.; Rocha, M., ProPythia: A Python package for protein classification based on machine and deep learning. *Neurocomputing* **2022**, *484*, 172-182.

(12) Merino, G. A.; Saidi, R.; Milone, D. H.; Stegmayer, G.; Martin, M. J., Hierarchical deep learning for predicting GO annotations by integrating protein knowledge. *Bioinformatics* **2022**, *38* (19), 4488-4496.

(13) Pan, T.; Li, C.; Bi, Y.; Wang, Z.; Gasser, R. B.; Purcell, A. W.; Akutsu, T.; Webb, G. I.; Imoto, S.; Song, J., PFresGO: an attention mechanism-based deep-learning approach for protein annotation by integrating gene ontology inter-relationships. *Bioinformatics* **2023**, *39* (3), btad094.

(14) Sanderson, T.; Bileschi, M. L.; Belanger, D.; Colwell, L. J., ProteInfer, deep neural networks for protein functional inference. *eLife* **2023**, *12*, e80942.

(15) Yu, T.; Cui, H.; Li, J. C.; Luo, Y.; Jiang, G.; Zhao, H., Enzyme function prediction using contrastive learning. *Science* **2023**, *379* (6639), 1358-1363.

(16) Watanabe, N.; Yamamoto, M.; Murata, M.; Kuriya, Y.; Araki, M., EnzymeNet: residual neural networks model for Enzyme Commission number prediction. *Bioinform. Adv.* **2023**, *3* (1), vbad173.

(17) Tamir, A.; Salem, M.; Yuan, J. S., ProtEC: A Transformer Based Deep Learning System for Accurate Annotation of Enzyme Commission Numbers. *IEEE/ACM TCBB* **2023**, *20* (6), 3691-3702.

(18) Yuan, Q.; Xie, J.; Xie, J.; Zhao, H.; Yang, Y., Fast and accurate protein function prediction from sequence through pretrained language model and homology-based label diffusion. *Brief. Bioinform.* **2023**, *24* (3), bbad117.

(19) Boulahrouf, K.; Aliouane, S. E.; Chehili, H.; Skander Daas, M.; Belbekri, A.; Hamidechi, M. A., EZYDeep: A Deep Learning Tool for Enzyme Function Prediction based on Sequence Information. *Open Bioinform. J.* **2023**, *16*, e18750362230627.

(20) Fernández, D.; Olivera-Nappa, Á.; Uribe-Paredes, R.; Medina-Ortiz, D. Exploring Machine Learning Algorithms and Protein Language Models Strategies to Develop Enzyme Classification Systems. In Bioinformatics and Biomedical Engineering; Rojas, I., Valenzuela, O., Rojas Ruiz, F., Herrera, L. J., Ortuño, F., Eds.; Springer Nature Switzerland: Cham, **2023**; pp 307-319.

(21) Buton, N.; Coste, F.; Le Cunff, Y., Predicting enzymatic function of protein sequences with attention. *Bioinformatics* **2023**, *39* (10), btad620.

(22) Derry, A.; Altman, R. B., Explainable protein function annotation using local structure embeddings. *BioRxiv* **2023**, 2023.10.13.562298.

(23) Kim, G. B.; Kim, J. Y.; Lee, J. A.; Norsigian, C. J.; Palsson, B. O.; Lee, S. Y., Functional annotation of enzyme-encoding genes using deep learning with transformer layers. *Nat. Commun.* **2023**, *14* (1), 7370.

(24) Shi, Z.; Deng, R.; Yuan, Q.; Mao, Z.; Wang, R.; Li, H.; Liao, X.; Ma, H., Enzyme Commission Number Prediction and Benchmarking with Hierarchical Dual-core Multitask Learning Framework. *Research* **2023**, *6*, 0153.

(25) Ayres, G.; Munsamy, G.; Heinzinger, M.; Ferruz, N.; Yang, K.; Lorenz, P. HiFi-NN annotates the microbial dark matter with Enzyme Commission numbers. *Machine Learning for Structural Biology Workshop*, *NeurIPS* **2023**.

(26) Han, S. R.; Park, M.; Kosaraju, S.; Lee, J.; Lee, H.; Lee, J. H.; Oh, T. J.; Kang, M., Evidential deep learning for trustworthy prediction of enzyme commission number. *Brief. Bioinform.* **2023**, *25* (1), bbad401.

(27) Tang, W.; Deng, Z.; Zhou, H.; Zhang, W.; Hu, F.; Choi, K. S.; Wang, S., MVDINET: A Novel Multi-Level Enzyme Function Predictor With Multi-View Deep Interactive Learning. *IEEE/ACM TCBB* **2024**, *21* (1), 84-94.

(28) Pertusi, D. A.; Moura, M. E.; Jeffryes, J. G.; Prabhu, S.; Walters Biggs, B.; Tyo, K. E. J., Predicting novel substrates for enzymes with minimal experimental effort with active learning. *Metab. Eng.* **2017**, *44*, 171-181.

(29) Yang, M.; Fehl, C.; Lees, K. V.; Lim, E. K.; Offen, W. A.; Davies, G. J.; Bowles, D. J.; Davidson, M. G.; Roberts, S. J.; Davis, B. G., Functional and informatics analysis enables glycosyltransferase activity prediction. *Nat. Chem. Biol.* **2018**, *14* (12), 1109-1117.

(30) Mou, Z.; Eakes, J.; Cooper, C. J.; Foster, C. M.; Standaert, R. F.; Podar, M.; Doktycz, M. J.; Parks, J. M., Machine learning-based prediction of enzyme substrate scope: Application to bacterial nitrilases. *Proteins* **2021**, *89* (3), 336-347.

(31) Robinson, S. L.; Smith, M. D.; Richman, J. E.; Aukema, K. G.; Wackett, L. P., Machine learning-based prediction of activity and substrate specificity for OleA enzymes in the thiolase superfamily. *Synth. Biol.* **2020**, *5* (1), ysaa004.

(32) Robinson, S. L.; Terlouw, B. R.; Smith, M. D.; Pidot, S. J.; Stinear, T. P.; Medema, M. H.; Wackett, L. P., Global analysis of adenylate-forming enzymes reveals β-lactone biosynthesis pathway in pathogenic *Nocardia*. *J. Biol. Chem.* **2020**, *295* (44), 14826-14839.

(33) Banerjee, D.; Jindra, M. A.; Linot, A. J.; Pfleger, B. F.; Maranas, C. D., EnZymClass: Substrate specificity prediction tool of plant acyl-ACP thioesterases based on ensemble learning. *CRBIOT* **2022**, *4*, 1-9.

(34) Kroll, A.; Ranjan, S.; Lercher, M. J., A multimodal Transformer Network for protein-small molecule interactions enhances predictions of kinase inhibition and enzyme-substrate relationships. *PLoS Comput. Biol.* **2024**, *20* (5), e1012100.

(35) Kroll, A.; Ranjan, S.; Engqvist, M. K. M.; Lercher, M. J., A general model to predict small molecule substrates of enzymes based on machine and deep learning. *Nat. Commun.* **2023**, *14* (1), 2787.

(36) Lu, C.; Lubin, J. H.; Sarma, V. V.; Stentz, S. Z.; Wang, G.; Wang, S.; Khare, S. D., Prediction and design of protease enzyme specificity using a structure-aware graph convolutional network. *PNAS* **2023**, *120* (39), e2303590120.

(37) Rappoport, D.; Jinich, A., Enzyme Substrate Prediction from Three-Dimensional Feature Representations Using Space-Filling Curves. *J. Chem. Inf. Model.* **2023**, *63* (5), 1637-1648.

(38) Harding-Larsen, D.; Madsen, C. D.; Teze, D.; Kittilä, T.; Langhorn, M. R.; Gharabli, H.; Hobusch, M.; Otalvaro, F. M.; Kırtel, O.; Bidart, G. N.; Mazurenko, S.; Travnik, E.; Welner, D. H., GASP: A Pan-Specific Predictor of Family 1 Glycosyltransferase Acceptor Specificity Enabled by a Pipeline for Substrate Feature Generation and Large-Scale Experimental Screening. *ACS omega* **2024**, *9* (25), 27278-27288.

(39) Mellor, J.; Grigoras, I.; Carbonell, P.; Faulon, J. L., Semisupervised Gaussian Process for Automated Enzyme Search. *ACS Synth. Biol.* **2016**, *5* (6), 518-528.

(40) Heckmann, D.; Lloyd, C. J.; Mih, N.; Ha, Y.; Zielinski, D. C.; Haiman, Z. B.; Desouki, A. A.; Lercher, M. J.; Palsson, B. O., Machine learning applied to enzyme turnover numbers reveals protein structural correlates and improves metabolic models. *Nat. Commun.* **2018**, *9* (1), 5252.

(41) Kroll, A.; Engqvist, M. K. M.; Heckmann, D.; Lercher, M. J., Deep learning allows genome-scale prediction of Michaelis constants from structural features. *PLoS Biol.* **2021**, *19* (10), e3001402.

(42) Li, F.; Yuan, L.; Lu, H.; Li, G.; Chen, Y.; Engqvist, M. K. M.; Kerkhoven, E. J.; Nielsen, J., Deep learning-based $k_{cat}$ prediction enables improved enzyme-constrained model reconstruction. *Nat. Catal.* **2022**, *5* (8), 662-672.

(43) Kroll, A.; Rousset, Y.; Hu, X. P.; Liebrand, N. A.; Lercher, M. J., Turnover number predictions for kinetically uncharacterized enzymes using machine and deep learning. *Nat. Commun.* **2023**, *14* (1), 4139.

(44) Yu, H.; Deng, H.; He, J.; Keasling, J. D.; Luo, X., UniKP: a unified framework for the prediction of enzyme kinetic parameters. *Nat. Commun.* **2023**, *14* (1), 8211.

(45) Du, B. X.; Yu, H.; Zhu, B.; Long, Y.; Wu, M.; Shi, J. Y. GELKcat: An Integration Learning of Substrate Graph with Enzyme Embedding for $K_{cat}$ Prediction. *2023 IEEE BIBM*, **2023**, 408-411.

(46) Wang, T.; Xiang, G.; He, S.; Su, L.; Wang, Y.; Yan, X.; Lu, H., DeepEnzyme: a robust deep learning model for improved enzyme turnover number prediction by utilizing features of protein 3D-structures. *Brief. Bioinform.* **2024**, *25* (5), bbae409.

(47) Qiu, S.; Zhao, S.; Yang, A., DLTKcat: deep learning-based prediction of temperature-dependent enzyme turnover rates. *Brief. Bioinform.* **2023**, *25* (1), bbad506.

(48) Maranas, C.; Boorla, V. S., CatPred: A comprehensive framework for deep learning in vitro enzyme kinetic parameters kcat, Km and Ki. *BioRxiv* **2024**, 2024.03.10.584340.

(49) Wang, J.; Yang, Z.; Chen, C.; Yao, G.; Wan, X.; Bao, S.; Ding, J.; Wang, L.; Jiang, H., MPEK: a multitask deep learning framework based on pretrained language models for enzymatic reaction kinetic parameters prediction. *Brief. Bioinform.* **2024**, *25* (5), bbae387.

(50) Chakravorty, D.; Khan, M. F.; Patra, S.; equally, f., Thermostability of Proteins Revisited Through Machine Learning Methodologies: From Nucleotide Sequence to Structure. *Curr. Biotechnol.* **2017**, *6*, 39-49.

(51) Rezaeenour, J.; Yari Eili, M.; Roozbahani, Z.; Ebrahimi, M., Prediction of Protein Thermostability by an Efficient Neural Network Approach. *Health Man. & Info. Sci.* **2016**, *3* (4), 102-110.

(52) Li, G.; Buric, F.; Zrimec, J.; Viknander, S.; Nielsen, J.; Zelezniak, A.; Engqvist, M. K. M., Learning deep representations of enzyme thermal adaptation. *Protein Sci.* **2022**, *31* (12), e4480.

(53) Yu, C. H.; Khare, E.; Narayan, O. P.; Parker, R.; Kaplan, D. L.; Buehler, M. J., ColGen: An end-to-end deep learning model to predict thermal stability of de novo collagen sequences. *J. Mech. Behav. Biomed. Mater.* **2022**, *125*, 104921.

(54) Yang, Y.; Zhao, J.; Zeng, L.; Vihinen, M., ProTstab2 for Prediction of Protein Thermal Stabilities. *Int. J. Mol. Sci.* **2022**, *23* (18), 10798.

(55) Jung, F.; Frey, K.; Zimmer, D.; Mühlhaus, T., DeepSTABp: A Deep Learning Approach for the Prediction of Thermal Protein Stability. *Int. J. Mol. Sci.* **2023**, *24* (8), 7444.

(56) Li, M.; Wang, H.; Yang, Z.; Zhang, L.; Zhu, Y., DeepTM: A deep learning algorithm for prediction of melting temperature of thermophilic proteins directly from sequences. *Comput. Struct. Biotechnol. J.* **2023**, *21*, 5544-5560.

(57) Wang, L.; Li, C., Optimal subset selection of primary sequence features using the genetic algorithm for thermophilic proteins identification. *Biotechnol. Lett.* **2014**, *36* (10), 1963-1969.

(58) Fan, G. L.; Liu, Y. L.; Wang, H., Identification of thermophilic proteins by incorporating evolutionary and acid dissociation information into Chou's general pseudo amino acid composition. *J. Theor. Biol*. **2016**, *407*, 138-142.

(59) Tang, H.; Cao, R.-Z.; Wang, W.; Liu, T.-S.; Wang, L.-M.; He, C.-M., A two-step discriminated method to identify thermophilic proteins. *Int. J. Biomath.* **2016**, *10* (04), 1750050.

(60) Sharma, A.; Bagler, G.; Bera, D., Supervised learning of protein thermal stability using sequence mining and distribution statistics of network centrality. *BioRxiv* **2019**, 777177.

(61) Li, J.; Zhu, P.; Zou, Q. Prediction of Thermophilic Proteins Using Voting Algorithm. In Bioinformatics and Biomedical Engineering; Rojas, I., Valenzuela, O., Rojas, F., Ortuño, F., Eds.; Springer International Publishing: Cham, **2019**; pp 195-203.

(62) Feng, C.; Ma, Z.; Yang, D.; Li, X.; Zhang, J.; Li, Y., A Method for Prediction of Thermophilic Protein Based on Reduced Amino Acids and Mixed Features. *Front. Bioeng. Biotechnol.* **2020**, *8*, 285.

(63) Charoenkwan, P.; Chotpatiwetchkul, W.; Lee, V. S.; Nantasenamat, C.; Shoombuatong, W., A novel sequence-based predictor for identifying and characterizing thermophilic proteins using estimated propensity scores of dipeptides. *Sci. Rep.* **2021**, *11* (1), 23782.

(64) Foroozandeh Shahraki, M.; Farhadyar, K.; Kavousi, K.; Azarabad, M. H.; Boroomand, A.; Ariaeenejad, S.; Hosseini Salekdeh, G., A generalized machine-learning aided method for targeted identification of industrial enzymes from metagenome: A xylanase temperature dependence case study. *Biotechnol. Bioeng.* **2021**, *118* (2), 759-769.

(65) Meng, C.; Ju, Y.; Shi, H., TMPpred: A support vector machine-based thermophilic protein identifier. *Anal. Biochem.* **2022**, *645*, 114625.

(66) Ahmed, Z.; Zulfiqar, H.; Khan, A. A.; Gul, I.; Dao, F. Y.; Zhang, Z. Y.; Yu, X. L.; Tang, L., iThermo: A Sequence-Based Model for Identifying Thermophilic Proteins Using a Multi-Feature Fusion Strategy. *Front. Microbiol.* **2022**, *13*, 790063.

(67) Charoenkwan, P.; Schaduangrat, N.; Moni, M. A.; Lio, P.; Manavalan, B.; Shoombuatong, W., SAPPHIRE: A stacking-based ensemble learning framework for accurate prediction of thermophilic proteins. *Comput. Biol. Med.* **2022**, *146*, 105704.

(68) Erickson, E.; Gado, J. E.; Avilán, L.; Bratti, F.; Brizendine, R. K.; Cox, P. A.; Gill, R.; Graham, R.; Kim, D. J.; König, G.; Michener, W. E.; Poudel, S.; Ramirez, K. J.; Shakespeare, T. J.; Zahn, M.; Boyd, E. S.; Payne, C. M.; DuBois, J. L.; Pickford, A. R.; Beckham, G. T.; McGeehan, J. E., Sourcing thermotolerant poly(ethylene terephthalate) hydrolase scaffolds from natural diversity. *Nat. Commun.* **2022**, *13* (1), 7850.

(69) Pei, H.-L.; Li, J.; Ma, S. H.; Jiang, J.; Li, M.; Zou, Q.; Lv, Z. J. A. S., Identification of Thermophilic Proteins Based on Sequence-Based Bidirectional Representations from Transformer-Embedding Features. *Appl. Sci.* **2023**, 13(5), 2858.

(70) Haselbeck, F.; John, M.; Zhang, Y.; Pirnay, J.; Fuenzalida-Werner, J. P.; Costa, R. D.; Grimm, D. G., Superior protein thermophilicity prediction with protein language model embeddings. *NAR Genom. Bioinform.* **2023**, *5*(4), lqad087.

(71) Zhao, J.; Yan, W.; Yang, Y., DeepTP: A Deep Learning Model for Thermophilic Protein Prediction. *Int. J. Mol. Sci.* **2023**, *24* (3), 2217.

(72) Wan, H.; Zhang, Y.; Huang, S., Prediction of thermophilic protein using 2-D general series correlation pseudo amino acid features. *Methods* **2023**, *218*, 141-148.

(73) Xiang, X.; Gao, J.; Ding, Y., DeepPPThermo: A Deep Learning Framework for Predicting Protein Thermostability Combining Protein-Level and Amino Acid-Level Features. *J. Comput. Biol.* **2024**, *31* (2), 147-160.

(74) Yu, H.; Luo, X., ThermoFinder: A sequence-based thermophilic proteins prediction framework. *Int. J. Biol. Macromol.* **2024**, *270* (Pt 2), 132469.

(75) Zhang, G.; Li, H.; Fang, B., Discriminating acidic and alkaline enzymes using a random forest model with secondary structure amino acid composition. *Process Biochem.* **2009**, *44* (6), 654-660.

(76) Fan, G.-L.; Li, Q.-Z.; Zuo, Y.-C., Predicting acidic and alkaline enzymes by incorporating the average chemical shift and gene ontology informations into the general form of Chou's PseAAC. *Process Biochem.* **2013**, *48* (7), 1048-1053.

(77) Lin, H.; Chen, W.; Ding, H., AcalPred: a sequence-based tool for discriminating between acidic and alkaline enzymes. *PloS one* **2013**, *8* (10), e75726.

(78) Khan, Z. U.; Hayat, M.; Khan, M. A., Discrimination of acidic and alkaline enzyme using Chou's pseudo amino acid composition in conjunction with probabilistic neural network model. *J. Theor. Biol.* **2015**, *365*, 197-203.

(79) Wang, X.; Li, H.; Gao, P.; Liu, Y.; Zeng, W., Combining Support Vector Machine with Dual g-gap Dipeptides to Discriminate between Acidic and Alkaline Enzymes. *Lett. Org. Chem.* **2019**, *16* (4), 325-331.

(80) Khan, Z.; Pi, D.; Khan, I.; Nawaz, A.; Ahmad, J.; Hussain, M., piSAAC: Extended notion of SAAC feature selection novel method for discrimination of Enzymes model using different machine learning algorithm. *arXiv preprint arXiv*:2101.03126, **2020**.

(81) Chu, Y.; Yi, Z.; Zeng, R.; Zhang, G., Predicting the optimum temperature of β-agarase based on the relative solvent accessibility of amino acids. *J. Mol. Catal. B Enzym.* **2016**, *129*, 47-53.

(82) Yan, S.; Wu, G., Predictors for Predicting Temperature Optimum in Beta-Glucosidases. *J. Biomed. Sci. Eng.* **2019**, *12*, 414-426.

(83) Yan, S.; Wu, G., Predicting pH Optimum for Activity of Beta-Glucosidases. *J. Biomed. Sci. Eng.* **2019**, *12*(07), 354-367.

(84) Li, G.; Rabe, K. S.; Nielsen, J.; Engqvist, M. K. M., Machine Learning Applied to Predicting Microorganism Growth Temperatures and Enzyme Catalytic Optima. *ACS Synth. Biol.* **2019**, *8* (6), 1411-1420.

(85) Gado, J. E.; Beckham, G. T.; Payne, C. M., Improving Enzyme Optimum Temperature Prediction with Resampling Strategies and Ensemble Learning. *J. Chem. Inf. Model.* **2020**, *60* (8), 4098-4107.

(86) Shahraki, M. F.; Atanaki, F. F.; Ariaeenejad, S.; Ghaffari, M. R.; Norouzi-Beirami, M. H.; Maleki, M.; Salekdeh, G. H.; Kavousi, K., A computational learning paradigm to targeted discovery of biocatalysts from metagenomic data: A case study of lipase identification. *Biotechnol. Bioeng.* **2022**, *119* (4), 1115-1128.

(87) Zhang, Y.; Guan, F.; Xu, G.; Liu, X.; Zhang, Y.; Sun, J.; Yao, B.; Huang, H.; Wu, N.; Tian, J., A novel thermophilic chitinase directly mined from the marine metagenome using the deep learning tool Preoptem. *Bioresour. Bioprocess.* **2022**, *9*, 54.

(88) Gado, J. E.; Knotts, M.; Shaw, A. Y.; Marks, D.; Gauthier, N. P.; Sander, C.; Beckham, G. T., Machine learning prediction of enzyme optimum pH. *BioRxiv* **2024**, 2023.06.22.544776.

(89) Qiu, S.; Hu, B.; Zhao, J.; Xu, W.; Yang, A., Seq2Topt: a sequence-based deep learning predictor of enzyme optimal temperature. *BioRxiv* **2024**, 2024.08.12.607600.

(90) Zaretckii, M.; Buslaev, P.; Kozlovskii, I.; Morozov, A.; Popov, P., Approaching Optimal pH Enzyme Prediction with Large Language Models. *ACS Synth. Biol.* **2024**, *13* (9), 3013-3021.

(91) Agostini, F.; Cirillo, D.; Livi, C. M.; Delli Ponti, R.; Tartaglia, G. G., ccSOL omics: a webserver for solubility prediction of endogenous and heterologous expression in *Escherichia coli*. *Bioinformatics* **2014**, *30* (20), 2975-2977.

(92) Hebditch, M.; Carballo-Amador, M. A.; Charonis, S.; Curtis, R.; Warwicker, J., Protein-Sol: a web tool for predicting protein solubility from sequence. *Bioinformatics* **2017**, *33* (19), 3098-3100.

(93) Khurana, S.; Rawi, R.; Kunji, K.; Chuang, G. Y.; Bensmail, H.; Mall, R., DeepSol: a deep learning framework for sequence-based protein solubility prediction. *Bioinformatics* **2018**, *34* (15), 2605-2613.

(94) Rawi, R.; Mall, R.; Kunji, K.; Shen, C. H.; Kwong, P. D.; Chuang, G. Y., PaRSnIP: sequence-based protein solubility prediction using gradient boosting machine. *Bioinformatics* **2018**, *34* (7), 1092-1098.

(95) Mall, R., SolXplain: An Explainable Sequence-Based Protein Solubility Predictor. *BioRxiv* **2019**, 651067.

(96) Han, X.; Zhang, L.; Zhou, K.; Wang, X., ProGAN: Protein solubility generative adversarial nets for data augmentation in DNN framework. *Comput. Chem. Eng.* **2019**, *131*, 106533.

(97) Han, X.; Wang, X.; Zhou, K., Develop machine learning-based regression predictive models for engineering protein solubility. *Bioinformatics* **2019**, *35* (22), 4640-4646.

(98) Bhandari, B. K.; Gardner, P. P.; Lim, C. S., Solubility-Weighted Index: fast and accurate prediction of protein solubility. *Bioinformatics* **2020**, *36* (18), 4691-4698.

(99) Raimondi, D.; Orlando, G.; Fariselli, P.; Moreau, Y., Insight into the protein solubility driving forces with neural attention. *PLoS Comput. Biol.* **2020**, *16* (4), e1007722.

(100) Hou, Q.; Kwasigroch, J. M.; Rooman, M.; Pucci, F., SOLart: a structure-based method to predict protein solubility and aggregation. *Bioinformatics* **2020**, *36* (5), 1445-1452.

(101) Martiny, H. M.; Armenteros, J. J. A.; Johansen, A. R.; Salomon, J.; Nielsen, H., Deep protein representations enable recombinant protein expression prediction. *Comput. Biol. Chem.* **2021**, *95*, 107596.

(102) Wu, X.; Yu, L., EPSOL: sequence-based protein solubility prediction using multidimensional embedding. *Bioinformatics* **2021**, *37* (23), 4314-4320.

(103) Hon, J.; Marusiak, M.; Martinek, T.; Kunka, A.; Zendulka, J.; Bednar, D.; Damborsky, J., SoluProt: prediction of soluble protein expression in *Escherichia coli*. *Bioinformatics* **2021**, *37* (1), 23-28.

(104) Chen, J.; Zheng, S.; Zhao, H.; Yang, Y., Structure-aware protein solubility prediction from sequence through graph convolutional network and predicted contact map. *J. Cheminform.* **2021**, *13* (1), 7.

(105) Madani, M.; Lin, K.; Tarakanova, A., DSResSol: A Sequence-Based Solubility Predictor Created with Dilated Squeeze Excitation Residual Networks. *Int. J. Mol. Sci.* **2021**, *22* (24), 13555

(106) Thumuluri, V.; Martiny, H. M.; Almagro Armenteros, J. J.; Salomon, J.; Nielsen, H.; Johansen, A. R., NetSolP: predicting protein solubility in *Escherichia coli* using language models. *Bioinformatics* **2022**, *38* (4), 941-946.

(107) Chen, L.; Wu, R.; Zhou, F.; Zhang, H.; Liu, J. K., HybridGCN for protein solubility prediction with adaptive weighting of multiple features. *J. Cheminform.* **2023**, *15* (1), 118.

(108) Chen, J.; Qian, Y.; Huang, Z.; Xiao, X.; Deng, L., Enhancing Protein Solubility Prediction through Pre-trained Language Models and Graph Convolutional Neural Networks. *2023 IEEE BIBM*, **2023**, 11-16.

(109) Wang, C.; Zou, Q., Prediction of protein solubility based on sequence physicochemical patterns and distributed representation information with DeepSoluE. *BMC Biol.* **2023**, 21(1), 12.

(110) Wang, X.; Liu, Y.; Du, Z.; Zhu, M.; Kaushik, A. C.; Jiang, X.; Wei, D., Prediction of Protein Solubility Based on Sequence Feature Fusion and DDcCNN. *Interdiscip. Sci. Comput. Life Sci.* **2021**, *13* (4), 703-716.

(111) Mehmood, F.; Arshad, S.; Shoaib, M., RPPSP: A Robust and Precise Protein Solubility Predictor by Utilizing Novel Protein Sequence Encoder. *IEEE Access* **2023**, *11*, 59397-59416.